\documentclass[lettersize,journal]{IEEEtran}

\pdfoutput=1%For ARXIV

\IEEEoverridecommandlockouts

\usepackage{amsfonts}
\usepackage[dvips]{graphicx}
\usepackage{times}
\usepackage{cite}
\usepackage{amsmath}
\usepackage{cases}
\usepackage{array}
\usepackage{dsfont}
\usepackage{amssymb}

\usepackage{stfloats}

\usepackage{graphicx}
\usepackage{subfigure}
\usepackage{pdfpages}
\usepackage{footnote}
\usepackage{booktabs}
\usepackage{soul}
\usepackage{array}
\usepackage{multirow}
\usepackage{bm}
\usepackage{empheq}
\usepackage{amsthm}
\usepackage{algorithm}
\usepackage{algorithmic}
\usepackage{color}
\usepackage{diagbox}
\usepackage{caption}
\theoremstyle{plain}

\theoremstyle{definition}

\newtheorem{definition}{Definition}

\usepackage{cite}
\usepackage{amsmath,amssymb,amsfonts}
\usepackage{algorithmic}
\usepackage{graphicx}
\usepackage{textcomp}
\usepackage{xcolor}

\ifCLASSINFOpdf
\else
\fi

\begin{document}
\title{Deep Learning-Enabled Semantic Communication Systems with Task-Unaware Transmitter and Dynamic Data}

\vspace{-25pt}\author{\IEEEauthorblockN{$\text{Hongwei Zhang}$,~\IEEEmembership{Student Member,~IEEE}, $\text{Shuo Shao}$,~\IEEEmembership{Member,~IEEE,} $\text{Meixia Tao},$~\IEEEmembership{Fellow,~IEEE}, $\text{Xiaoyan Bi}$, and $\text{Khaled B. Letaief},~\IEEEmembership{Fellow,~IEEE}$
}
\vspace{-18pt}
\thanks{(\textit{Corresponding authors:Meixia Tao, Shuo Shao.})}
\thanks{Hongwei Zhang, Shuo Shao, and Meixia Tao are with the School of Electronic Information and Electric Engineering, 
Shanghai Jiao Tong University, Shanghai, China (e-mails: \{zhanghwei, shuoshao, mxtao\}@sjtu.edu.cn). Xiaoyan Bi is with the Ottawa research center, Huawei, Ottawa, Canada (e-mail: bixiaoyan@huawei.com). Khaled B. Letaief is with the Electronic and Computer Engineering Department, Hong Kong University of Science and Technology, Hong Kong, China (e-mail: eekhaled@ust.hk).}
\thanks{This work is supported by the National Key R\&D Project of China under grant 2019YFB1802702, the NSF of China under grants 62125108, 61901261, and 12031011.}
\thanks{The codes of the proposed method are available on \textit{https://github.com/SJTU-mxtao/Semantic-Communication-Systems.}}}

% make the title area
\maketitle

\begin{abstract}
Existing deep learning-enabled semantic communication systems often rely on shared background knowledge between the transmitter and receiver that includes empirical data and their associated semantic information. In practice, the semantic information is defined by the pragmatic task of the receiver and cannot be known to the transmitter. The actual observable data at the transmitter can also have non-identical distribution with the empirical data in the shared background knowledge library. To address these practical issues, this paper proposes a new neural network-based semantic communication system for image transmission, where the task is unaware at the transmitter and the data environment is dynamic. The system consists of two main parts, namely the semantic coding (SC) network and the data adaptation (DA) network. The SC network learns how to extract and transmit the semantic information using a receiver-leading training process. By using the domain adaptation technique from transfer learning, the DA network learns how to convert the data observed into a similar form of the empirical data that the SC network can process without re-training. Numerical experiments show that the proposed method can be adaptive to observable datasets while keeping high performance in terms of both data recovery and task execution.

\end{abstract}


\IEEEpeerreviewmaketitle

\section{Introduction} \label{Introduction}
\IEEEPARstart{W}{ith} the booming development of machine learning (ML) and computer hardware, native intelligence is envisioned to be an essential component of 6G mobile networks \cite{letaief2021edge}. As a new intelligence-enabled communication paradigm, semantic communication, sometimes also called task-oriented communication, has drawn lots of research interest recently. By extracting and transmitting the information that can best support the pragmatic task of the receiver, called \textit{semantic information} \cite{xie2021task}, semantic communication can significantly improve transmission efficiency and reliability \cite{lan2021semantic,zhang2021toward}. It is promising for a wide range of human-centric and machine-centric applications, such as smart transportation, augmented reality (AR), interactive hologram, and intelligent humanoid robots. 

Based on the information-theoretic formulation of semantic communication in \cite{bao2011towards,9518240}, a number of deep learning (DL)-enabled semantic communication systems have been proposed \cite{liew2021economics,lan2021semantic,liu2021semantics,yang2021semantic,shi2021knowledge,weng2021semantic1,xie2021task,xie2020deep}. Despite the difference in details, these systems share a common general framework. Specifically, semantic information is regarded as the hidden features of observable information, where the relationship between semantic information and observable information is not directly given. However, empirical data tuple of observable information and associated semantic information is provided in a large enough knowledge base. Since the joint distribution between semantic information and observable information could be too complicated to accurately estimate by traditional methods, neural networks (NNs) are hence deployed. They will be trained with the knowledge base, learning how to extract the semantic information and transmit it over communication channels \cite{liu2021semantics,yang2021semantic,shi2021knowledge}.
%\textcolor{red}{[the details of these work are presented in Section II.]}. 
%For example, convolutional neural networks (CNN) is deployed as the semantic encoder for graphic and speech data transmission \cite{yang2021semantic,weng2021semantic1}. 
%\tao{"should you mention what is semantic coding somewhere around here?"} 
This NN-based semantic communication framework is capable of processing various types of data, including images \cite{yang2021semantic}, vocal speeches \cite{weng2021semantic1}, language texts \cite{xie2021task}, etc.
Compared to communication systems with traditional source and channel coding algorithms applied, these NN-based semantic communications can largely improve the effectiveness of pragmatic task execution such as classification, detection, and other image processing tasks, under the same transmitting rate. 

%In more general and practical scenarios, the existence of such a complete and precise knowledge base is very ideal setup that is too hard to achieve. At first, establishing a proper knowledge base is still a difficult problem in statistics. Since evaluating the variability between the semantic knowledge base and to-be-transmitted data is an open problem \cite{shi2021semantic}, it cannot be ensured that the knowledge base can precisely represent the to-be-transmitted data. Second, even if there is a representative set of empirical data, the semantic information can still be hardly known by the transmitter in practical scenarios. As pointed out in a recent paper \cite{xie2021task}, the semantic information should be defined by the pragmatic task of the receiver, instead of an intrinsic property of the observable information.  

However, it is worthwhile to mention that in many practical scenarios, the original setup of fully shared background knowledge with complete observable and semantic information does not stand for granted. As pointed out in \cite{xie2021task}, semantic information is the information that can best support the pragmatic task of the receiver, which makes it task-oriented information instead of an intrinsic property of the observable information. For example, the pragmatic task on an image might be object recognition, target detection and etc., and the corresponding semantic information for these tasks is also different. Since the pragmatic task is typically unknown to the transmitter, the implementation of the training process at the transmitter and the receiver, which has been overlooked in many existing works, is non-trivial. Ideally, the receiver can feed all of its local empirical semantic information that is relevant to its personalized tasks back to the transmitter. But this procedure will have a high communication and time cost. It may also suffer privacy concerns if the receiver does not want the transmitter to know about its own pragmatic use of the data.

Meanwhile, as considered in \cite{xie2020deep}, the distribution of the transmitted data can be varying and different from that of the original background knowledge. For example, the training data may be drawn from the MNIST dataset, but the actual transmit data may be drawn from the SVHN dataset, which has a similar but different distribution from the previous one. As the scalability of NN is limited, the performance of NN will decline when the data distributions are changed. This is similar to the over-fitting phenomenon when the assumptions of training samples are different from those of test samples. Re-training NN can improve performance, but again, it requires extra communication and computing overhead.

%Motivated by previous works, in this paper we focus on a speciachangedl semantic communication system, where the transmitter is unaware of the pragmatic task, and the empirical data and future data may be non-identically distributed. In this context, task-unaware means neither the data set to transmit (which hereby we call as ``target data" for short), nor the pragmatic task on this dataset, is known to the transmitter. This setup leads to two major challenges. First, the transmitter has to learn what the semantic information is before learning how to extract the semantic information. Meanwhile, the communication cost in this learning process should be considered. Second, the semantic coding strategy should be adaptive to different data sets. Of course, ideally we can retrain the whole network every time when a new data set appears. But considering the necessary data for neural network training is absent at the receiver end, this procedure will have an intolerably high communication and time cost, which makes it impractical. 

To address the above practical issues, in this paper, we focus on a receiver-leading dynamic semantic communication system, for image transmission in particular, where the transmitter is unaware of the pragmatic task. Moreover, the neural network training in this system is divided into two stages, namely the preparation stage and the working stage, where the actual observable data in the working stage has non-identical distribution from the empirical data that is known in the working stage as background knowledge. 
This setup leads to two major challenges in the NN-based approach. First, the receiver has to enlighten the transmitter on how to encode semantic information and observable information in the training process, since the receiver-specific empirical semantic information is unknown to the transmitter and cannot be sent to the transmitter either. Second, the semantic communication strategy should be adaptive to accommodate the possible difference between empirical data and observable data. A transfer learning-based method is applied to accelerate the model re-training in \cite{xie2020deep}, however, it still requires extra communication between transmitter and receiver for the transfer learning of the coder NNs.

In our considered semantic communication system with the task-unaware transmitter in a dynamic data environment, we propose a new neural network-based semantic communication framework to solve the two challenges above. There are two goals in this semantic communication system, i.e., pragmatic task \cite{yang2021semantic,dai2021semantic} and observable information reconstruction \cite{shi2021knowledge,wang2022wireless}. The proposed framework consists of a semantic coding network and a data adaptation network, which are trained in two separated stages. More specifically, in the first stage the semantic coding network is jointly trained at the transmitter and receiver to encode observable information and decode accordingly based on the empirical data in background knowledge whose semantic information is known to the receiver only. In the second stage, the data adaptation network is trained at the transmitter side only to convert the newly observed type of data into the form of data that the semantic coding network trained in the first stage can be re-used without re-training. 

The main contribution of this paper is three-folded. First, a receiver-leading training process, which is not considered in most existing works, is proposed for the semantic coding network with the task-unaware transmitter. With this process, the receiver can coordinate the network training at the transmitter, without announcing to the transmitter what the task exactly is. Hence the transmitter can learn how to encode the observable information with some limited feedback from the receiver. As such, the focus is on the design of crucial feedback information for network training at the transmitter without disclosing the specific task carried out at the receiver, rather than the design of specific NN architecture as considered in the literature. Most existing works suppose that the knowledge of the task is available to both the transmitter and the receiver, they thus do not consider the training process of semantic communication networks.

Second, a set of loss functions tailored for image transmission are proposed for this semantic coding network accordingly. Though the structure of loss functions that consists of bit-wised distortion measure and semantic-wised distortion measure has already been proposed in some previous works \cite{liu2021semantics,shi2021new}, we sort the tasks into two categories based on the output of the pragmatic function and propose the distortion measure function for each category especially. Moreover, an algorithm in the empirical sampling form for NN training is also proposed accordingly, in order to fit the theoretic loss function in probabilistic form into our training process.

The third contribution is the introduction of the data adaptation network at the transmitter to tackle the issue with different data environments. By using the domain adaptation technique of transfer learning, this network can be trained locally at the transmitter end without any communication to the receiver needed, and with only a few shots of observable data. Generally, the more similar the historical empirical data is to the current observable data, the more efficient our data adaptation network will be. Hence, we also propose a theoretic measure function to sketch this similarity between datasets.

Extensive numerical experiments are conducted in this paper to verify the effectiveness of the proposed semantic communication system. First, it is demonstrated that the proposed receiver-leading training process can well train the semantic coding network for the receiver-specific pragmatic tasks, including handwritten digit recognition, image classification, and image segmentation, with no direct information on pragmatic tasks disclosed to the transmitter. It is also shown that, with the proposed loss function design, the semantic coding network is able to strike a flexible balance between data recovery performance, e.g. peak signal-to-noise ratio (PSNR), and task execution performance, e.g., classification accuracy, in response to different compression rates, as compared with existing schemes. The proposed scheme outperforms baseline schemes, especially when the source is complex and the channel condition is poor. Finally, when the observable dataset is different from the library dataset, experimental results show that the proposed data adaptation method without re-training the semantic coding network significantly outperforms the case if no data adaptation is employed. Meanwhile, it also has a very close performance to the ideal case with full retraining. 

The rest of the paper is organized as follows. In Section \ref{sec:related_work}, related work is presented. In Section \ref{sec:system}, the formal problem setup will be introduced. In Sections \ref{sec:coding} and \ref{sec:DA}, we will show the semantic coding network training process and the data adaptation network design, respectively. In Section \ref{sec:results}, experimental results will be given. Finally, Section \ref{sec:conclusion} will conclude the paper.

\section{Related Work}
\label{sec:related_work}
\subsection{Definition of Semantic Information}
In \cite{carnap1952outline}, R. Carnap \textit{et al.} first discovered a problem with Shannon's information theory, namely that a self-contradictory sentence is supposed to contain a great deal of information in Shannon's information theory, yet the receiver will not accept it and think this sentence has no information. Inspired by this, a theory of strong semantic information based on truth value rather than probability distribution was proposed \cite{floridi2004outline,bao2011towards}. Semantic information is also defined as the syntactic information that a physical system has about its environment in \cite{kolchinsky2018semantic}, and as the relative importance of random information sources in \cite{kountouris2021semantics}. In addition, there are many other works that have defined semantic information from different aspects, as described in \cite{resnik1995using,guler2018semantic,chaitin1977algorithmic,zhong2017theory}. 

\subsection{NN-based Semantic Communications}
With the development of artificial intelligence (AI), an increasing number of AI-enabled semantic communication systems have been proposed, which can automatically learn to extract and transmit semantic information based on the data distribution, task, and channel status. Specifically, the framework of semantic communication between intelligent agents is proposed in \cite{shi2021new}. A multi-semantic communication system for the image classification task with different granularity is built based on a fully convolutional network in \cite{liu2021semantics}. A multi-modal semantic communication system based on long short term memory (LSTM) \cite{van2020review} is proposed in \cite{xie2021task}, which can transmit both image and text simultaneously. Besides, a multi-user semantic system is proposed, where partial users transmit images while others transmit texts to inquire about the information about the images \cite{xie2021task}. A semantic signal processing framework based on ResNet \cite{targ2016resnet}, which can be changed between specific tasks easily, is proposed in \cite{kalfa2021towards}. In the Internet of Things (IoT), where the computing power of each device is limited, a semantic communication system based on ADNet \cite{tian2020attention} is proposed to transmit text in \cite{xie2020lite}. 

Given the above, one can see that AI-enabled methods can train a variety of models for transmitting various kinds of data, such as image \cite{liu2021semantics,hu2022robust}, video \cite{jiang2022wireless}, speech \cite{weng2021semantic,weng2021semantic1} and text \cite{xie2021deep,tung2021deepwive}. These previous works utilize AI methods to jointly train neural networks (NNs) as semantic encoders and decoders to recover the semantic information, in terms of some given measure such as Kullback-Leibler (KL) divergence or quadratic loss \cite{liu2021semantics,shi2021knowledge}. This method has proved to be effective in many practical cases when the semantic information is labeled in advance as ground truth \cite{liew2021economics,yang2021semantic}.  

\subsection{Previous Works on Domain Adaptation}
Another important area touched in this paper is domain adaptation (DA). As a special kind of transfer learning \cite{pan2009survey}, DA fulfills new tasks by transforming the samples in the source domains into the samples in the related target domain \cite{ganin2016}. DA is commonly used in computer vision because the domain of the dataset for training and that for inference in the tasks are often different. As such, DA is naturally suitable to solve the mismatch issues between the actual observable dataset and the empirical dataset considered in this paper.

Recent advances on DA can be categorized as follows: 
\begin{itemize}
    \item DA based on divergence, which is implemented by minimizing the divergence between the data distribution in the source domain and the target domain to achieve domain-invariant feature representation \cite{rozantsev2018beyond,sun2016deep,damodaran2018deepjdot}.
    \item DA based on reconstruction, which not only can learn to correctly identify the target data, but can also save information about the target data \cite{ghifary2016deep,zhu2017unpaired,isola2017image}.
    \item Adversarial DA, which utilizes generative adversarial networks (GAN) to generate synthetic target data related to the source domain (for example by retaining labels) \cite{yoo2016pixel,bousmalis2017unsupervised,ganin2016}.
\end{itemize} 

Among the above DA methods, adversarial DA will be adopted in our work. Specifically, we will utilize the GAN to generate the data to ensure that the receiver can understand and execute the task accordingly.

\begin{figure*}[t]
\centering
\includegraphics[scale=0.21]{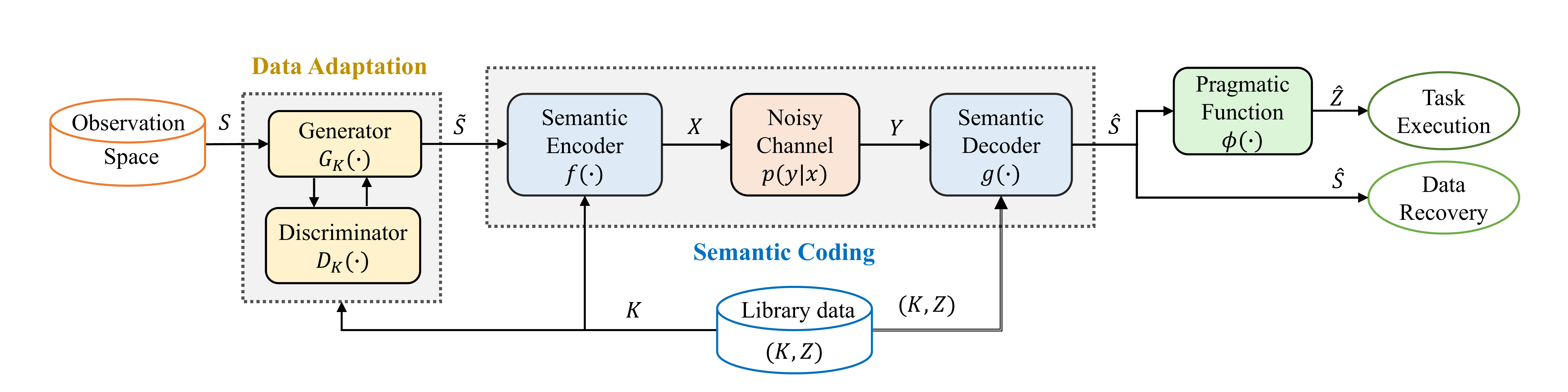}
\caption{Illustration of the main parts in the proposed semantic communication system.}
\vspace{-12pt}
\label{fig:system_all}
\end{figure*}

\section{System Model}
\label{sec:system}
In this section, we give the formal introduction of our proposed DL-enabled semantic communication system with a task-unaware transmitter in the dynamic data environment, including the physical communication model and the proposed learning model.

The overall system architecture is shown in Fig. \ref{fig:system_all}. 
There is a dataset called ``observation space" $\mathcal{S}$, whose empirical knowledge is unknown at the transmitter prior to system operation. There is also another dataset called ``library dataset" $\mathcal{K}$, which is a collection of empirical data $K$ and its corresponding pragmatic task $Z$ from background knowledge. Our system works in two stages, namely the preparation stage and the working stage. In the preparation stage, the transmitter and receiver train the coder networks collectively with the empirical data $K$. While in the working stage, the transmitter further trains the data adaptation network with both the empirical data $K$ and the observable data $S$. The alphabet of $S$, $K$ and $Z$ are denoted as $\Gamma_{\mathcal{S}}$, $\Gamma_{\mathcal{K}}$, and $\Gamma_{\mathcal{Z}}$, respectively. In this paper, we focus on image data in RGB format. Hence, both $S$ and $K$ are modeled as sequences of three-dimensional vectors with 8-bit symbols. As a remark, the empirical data $K$ is known by both the transmitter and receiver, while its pragmatic task $Z$ is only available to the receiver. Meanwhile, the empirical data $K$ and observable data $S$ may be drawn from different observation spaces. To our best knowledge, a knowledge base with different access levels has not been considered in the existing literature on semantic communications. Though semantic communications with a dynamic data environment have been studied in \cite{xie2020deep}, however, their method cannot be applied here due to the privacy requirement in our model.

The data communication process of this system is as follows. The goal of the transmitter is to send some data to the receiver for both pragmatic task use and observable data recovery. When the receiver initiates a data request, the transmitter will sample data $S$ from the observation space. If the observation space and the library set indeed follow different distributions, the sampled data $S$ will be converted into a different form of data, denoted by $\widetilde{S}$, by a function $G_{K}(\cdot)$ as $\widetilde{S}=G_{K}(S)$. The transferred data $\widetilde{S}$, which is in a similar form to the empirical data $K$, will be encoded by an encoder $f(\cdot)$ to get the channel input $X$, whose alphabet is denoted as $\Gamma_{\mathcal{X}}$. For the image source with RGB format, both the sampled data $S$ and the empirical data $K$ are modeled as a sequence of three-dimensional vectors with 8-bit symbols. In order to coincide with the RGB format image source, the channel input $X$ is also set as a sequence of the three-dimensional vectors with 8-bit symbols. As such, the \textit{compression rate} can be defined as $\textit{CR}=\log |\Gamma_{\mathcal{X}}|/\log |\Gamma_{\mathcal{K}}|$, which is equivalent to the ratio between image resolution and the sequence length of channel input. The encoded signal $X$ will be transmitted through an AWGN channel, where the channel output is denoted as $Y=X+N$. The power of channel noise $N$ is fixed but unknown.
Upon receiving $Y$, the receiver can reconstruct a distorted version of observable data as $\widehat{S}$ according to decoder $g(\cdot)$, and further reconstruct a distorted pragmatic task output $\widehat{Z}$ by pragmatic task function $\phi(\cdot)$.

In our approach, the data converting function, the encoding function, and the decoding functions, namely $G_{K}(\cdot)$, $f(\cdot)$, $g(\cdot)$, are all realized by neural networks. Meanwhile, in this paper, we set the pragmatic function $\phi(\cdot)$ as a given one. It can be regarded as the mathematical abstraction of the pragmatic task, while how to obtain it is not the focus of this paper.

As mentioned, the three networks are trained in two stages sequentially. The first stage is the preparation stage, which is before the appearance of observable data. In this stage, the encoder $f(\cdot)$ and the decoder $g(\cdot)$ are jointly trained, based on the library dataset $\cal K$. The joint source-channel coding (JSCC)-based semantic communication in this paper can also be regarded as task-oriented semantic communication \cite{gunduz2022beyond,2018game}. Hence, the encoder $f(\cdot)$ aims to extract and transmit the data containing most semantic information and observable information, while the decoder $g(\cdot)$ aims to reconstruct the data related to the pragmatic task and the empirical data. Also mentioned in \cite{shi2021semantic,xie2021task}, the coders take into account both the semantic information and channel influence.  Since the empirical data $K$ is known to both the transmitter and the receiver, but the empirical pragmatic task $Z$ is only available to the receiver, hence the receiver will lead the training process. By comparing the reconstructed data and raw data, the receiver can teach the transmitter how to improve its encoder. The detailed training process of the receiver-leading semantic coding network and its loss function design will be described later in Section \ref{sec:coding}.

The second stage is the working stage, which is after enough observable data $S$ is collected. In this stage, the function $G_{K}(\cdot)$ is trained to convert $S$ into a similar form of the empirical data $K$, so that the well-trained encoder $f(\cdot)$ can still extract and transmit the semantic information without re-training. The network of $G_{K}(\cdot)$ is called the data adaptation network. A discriminator is introduced to help this conversion. When the discriminator is unable to distinguish the converted data and the empirical data from the library set, the converter is regarded as good enough. In this training process, no information on the pragmatic task is required. Hence the second stage of training can be conducted locally at the transmitter end, without communication to the receiver. The detailed architecture of the data adaptation network and its loss function design will be described later in Section \ref{sec:DA}. \par

\begin{figure*}[t]
\centering
\includegraphics[scale=0.21]{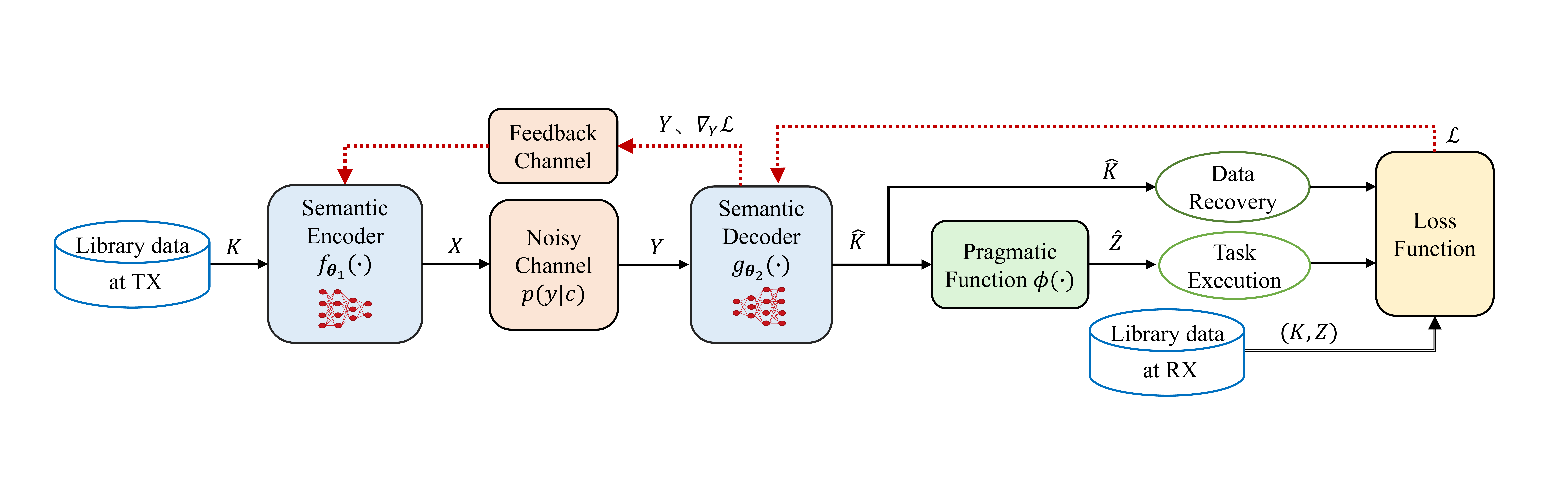}
\caption{Illustration of the semantic coding part in the proposed semantic communication system. }
\vspace{-12pt}
\label{fig:system_training}
\end{figure*}

\section{The Semantic Coding Network}
\label{sec:coding}
In this section, we give a detailed illustration of the semantic coding network, including the receiver-leading training process in Section \ref{sec:preliminary}, where the training strategy and algorithm are proposed; and the corresponding loss function design in Section \ref{objective function}, where both the theoretic loss function and explicit training function for each epoch are provided.
\par

\subsection{Semantic Coding Network Training}
\label{sec:preliminary}

\begin{algorithm}[t]
% \fontsize{9pt}{11pt}\selectfont
\fontsize{10pt}{11pt}\selectfont
\caption{semantic coding training algorithm.}
\label{preliminary_traning}
\begin{algorithmic}[1]

\STATE Set epoch counter $t=1$.

\WHILE{the training stop condition is not met}
\STATE Take a batch of the samples $\mathcal{K}_t \subset \mathcal{K}$ (transmitter)
\STATE Encode and send all $X= f_{\boldsymbol{\theta}_1,t}\left(K\right)$ in $\mathcal{K}_t$ (transmitter).
\STATE Decode data $ \widehat{K} = g_{\boldsymbol{\theta}_2,t}\left(Y\right)$ (receiver).
\STATE $\widehat{Z} = \phi\left(\widehat{K}\right) $ (receiver).
\STATE Calculate the gradients $ \nabla_{\boldsymbol{\theta}_2} \mathcal{L}(T) $ based on \eqref{eq:gra_decoder} and update $g_{\boldsymbol{\theta}_2}(\cdot)$ (receiver).
\STATE Calculate the gradient $ \nabla_{Y} \mathcal{L}$ (receiver).
\STATE Sends  $ \nabla_{Y} \mathcal{L} $ and $Y$ to the transmitter (receiver).
\STATE Calculate $\mathbb{E}_{\sim \mathcal{T}_t}\left[\nabla_{\boldsymbol{\theta}_1} \mathcal{L}\right]$ based on \eqref{eq:chain} and update $f_{\boldsymbol{\theta}_1}\left(\cdot\right)$ (transmitter).

\STATE $ t=t+1 $.
\ENDWHILE

\end{algorithmic}
\end{algorithm}

The system flow for semantic coding is shown in Fig. \ref{fig:system_training}. The semantic encoder $f(\cdot)$ aims to extract and transmit the information which can recover observable information and are most correlated with the pragmatic task $Z$, and the decoder $g(\cdot)$ aims to reconstruct the empirical data correspondingly. We model the encoder $f_{\boldsymbol{\theta}_1}(\cdot)$ and decoder $g_{\boldsymbol{\theta}_2}(\cdot)$ as two deep neural networks (DNNs), where $\boldsymbol{\theta}_1$ and $\boldsymbol{\theta}_2$ are two sets of network parameters, respectively. Here there is no specific requirement on the network architecture.

\begin{figure*}[!b]
\hrule
\begin{align}
\nabla_{\boldsymbol{\theta}_1} ( \mathcal{L}_{\boldsymbol{\theta}_1,\boldsymbol{\theta}_2}(K,\widehat{K},Z,\widehat{Z}) )
& = {\underbrace{\nabla_{Y}(\mathcal{L}_{\boldsymbol{\theta}_1,\boldsymbol{\theta}_2}(K,\widehat{K},Z,\widehat{Z}))}_{\text{at the receiver}}}\cdot {\underbrace{\nabla_{\boldsymbol{\theta}_1}Y}_{\text{at the transmitter}}}
\label{eq:chain} \\
\nabla_{\boldsymbol{\theta}_2}(\mathcal{L}_{\boldsymbol{\theta}_1,\boldsymbol{\theta}_2}(K,\widehat{K},Z,\widehat{Z})) &= \nabla_{\widehat{K}}(\mathcal{L}_{\boldsymbol{\theta}_1,\boldsymbol{\theta}_2}(K,\widehat{K},Z,\widehat{Z})) \cdot \nabla_{\boldsymbol{\theta}_2}\widehat{K}
\label{eq:gra_decoder}
\end{align}
% \hrule
\end{figure*}

The training process is as follows. In each training epoch $t \in \{1,2,\cdots\}$, the transmitter draws samples of $K$ from library dataset $\mathcal{K}$ uniformly at random to form a training batch $\mathcal{K}_t$. For each sample $K\in \mathcal{K}_t$, the transmitter encodes it as $X= f_{\boldsymbol{\theta}_1,t}\left(K\right)$ and then sends the encoded signal $X$ to the receiver. Hence the receiver can obtain reconstructed data $\widehat{K} = g_{\boldsymbol{\theta}_2,t}\left(Y\right)$ and the corresponding reconstructed pragmatic output $\widehat{Z} = \phi\left(\widehat{K}\right)$. Meanwhile, the receiver is also aware of the original sample $K$ and the ground-truth pragmatic output $Z$ of $K$. 
Hence, a complete training sample is defined as a tuple $T \triangleq (K,\widehat{K},Z,\widehat{Z})$, and a collection of these samples form a complete random training batch, which is denoted as $\mathcal{T}_t$. As a remark, all contents in the complete training sample $\mathcal{T}_t$ are fully available to the receiver, while the transmitter only has the data in training batch $\mathcal{K}_t$.

Our goal is to find the optimal encoding and decoding functions so that the loss of both semantic information and observable information can be minimized. Equivalently, it is to optimize the parameters $\boldsymbol{\theta}_1$ and $\boldsymbol{\theta}_2$. 
We call this loss function \textit{semantic distortion (SD)} and denote it as $\mathcal{L}(K,\widehat{K},Z,\widehat{Z})$, or $\mathcal{L}(T)$ for short. When the average gradients over the training batch denoted as $\mathbb{E}_{\sim \mathcal{T}_t}\left[\nabla_{\boldsymbol{\theta}_1}\mathcal{L}\right]$ and $\mathbb{E}_{\sim \mathcal{T}_t}\left[\nabla_{\boldsymbol{\theta}_2}\mathcal{L}\right]$ are known by the transmitter and receiver respectively, we can utilize the Adam algorithm \cite{Adam} to update the parameters as
\begin{align}
\boldsymbol{\theta}_{i,t+1}=\boldsymbol{\theta}_{i,t} - \eta \frac{\rho_t}{\sqrt{\nu_t+\epsilon}} \quad\quad (i\in \{1,2\}),
\label{eq:Adam}
% \vspace{-15pt}
\end{align}
where $ \rho_t $ and $ \nu_t $ are the first and second order momenta of gradients, respectively, $ \epsilon $ is a smooth term to prevent the denominator of the second term from being zero, and $\eta$ is the learning rate. \par

Next, we will present how to obtain $\mathbb{E}_{\sim \mathcal{T}_t}\left[\nabla_{\boldsymbol{\theta}_1}\mathcal{L}\right]$ and $\mathbb{E}_{\sim \mathcal{T}_t}\left[\nabla_{\boldsymbol{\theta}_2}\mathcal{L}\right]$, which is the essential issue. As mentioned, the complete training samples are available at the receiving end, hence $\nabla_{\boldsymbol{\theta}_2}\mathcal{L}(T)$ for each sample $T$ can be readily obtained by the receiver as \eqref{eq:gra_decoder}. However, the transmitter only has the information of $K$, so it requires the receiver to feedback some necessary values in order to obtain $\nabla_{\boldsymbol{\theta}_1}\mathcal{L}(T)$. According to the chain rule, the gradient of loss function over the encoder network parameter, namely $\nabla_{\boldsymbol{\theta}_1} \mathcal{L}(K,\widehat{K},Z,\widehat{Z}) $, can be derived as \eqref{eq:chain}. 
Specifically, the item in the first underbrace, i.e., the gradient of the loss function with respect to the channel output, can be calculated numerically at the receiver end, and then sent to the transmitter. Meanwhile, since $\boldsymbol{\theta}_1$ is unknown to the receiver, the receiver sends the channel output $Y$ back to the transmitter and lets the transmitter compute $\nabla_{\theta_1} Y$ locally, i.e., the second term in \eqref{eq:chain}. Hence the feedback content can be written as the data tuple $\{\nabla_{Y} \mathcal{L}(T),Y\}$. We can see that no direct information about the pragmatic function is leaked to the transmitter, which ensures a weak level of security. That is, knowing $\nabla_{Y} \mathcal{L}$ alone is not sufficient for the transmitter to determine the pragmatic use of its transmitted data. In addition, from the information-theoretic security perspective, though secrecy is a quantitative problem, our proposed feedback method is still secure enough to prevent privacy leakage when the size of the transmitted data is sufficiently larger than that of the pragmatic output as shown in \cite{ding2019submodularity}.  \par

Besides, since the transmitter and receiver are regarded as unequal participants in our model, where the receiver is the dominant one with more communication resources and data authorization privilege, we assume that the receiver has enough power to enjoy a noiseless feedback channel noiseless.

The above training procedure is outlined in Algorithm \ref{preliminary_traning}. Since the NNs are trained over epochs, here we define $f_{\boldsymbol{\theta}_1,t}\left(\cdot\right)$ and $g_{\boldsymbol{\theta}_2,t}\left(\cdot\right)$ as the encoding and decoding functions of the $t$-epoch, which will converge with the training process.
Note that the pragmatic function $\phi(\cdot)$ is assumed known by the receiver in advance, as we already mentioned in Section \ref{sec:system}.

\subsection{Loss Function Design for Semantic Coding} \label{objective function}

Compared with the traditional coding system, an essential difference of semantic communication is that not only the performance of traditional data recovery but also the performance for further pragmatic use are required.
In the work of Liu \textit{et. al.} \cite{9518240}, the semantic source coding is modeled as a rate-distortion problem, where there are two distortion constraints for recovering semantic information $S$ and observable information $X$ respectively. 
Many other existing works on NN-based semantic communication systems \cite{liu2021semantics,shi2021new} also follow this idea. The loss function of these neural networks all contains both items of semantic information loss and observable information loss, in the KL divergence or quadratic measure of information difference. In the considered task-unaware semantic communication system, we inherit this binary structure with both semantic information and observable information.

With the information of $\{\nabla_{Y} \mathcal{L}(T),Y\}$ sent back from the receiver, the transmitter can adjust its encoding neural network accordingly. That is to say, the transmitter can learn the encoding function that contains the most semantic information, without exactly knowing what the semantic information is. Next, we will define semantic distortion mathematically.

\begin{definition}
\label{def1}
Consider the scenario where the semantic encoder and decoder are jointly optimized. The loss function is defined as \textit{semantic distortion (SD)}, which is expressed as
\begin{equation}
\mathcal{L}_{\boldsymbol{\theta}_1,\boldsymbol{\theta}_2}(K, \widehat{K}, Z, \widehat{Z})  \triangleq \! \lambda \alpha \mathcal{D}_{\rm ob}(K,\widehat{K}) + (1-\lambda)\mathcal{D}_{\rm pr}(Z,\widehat{Z}), \label{eq:coding}
\end{equation}
where $\mathcal{D}_{\rm ob}$ and $\mathcal{D}_{\rm pr}$ are distortion measure functions for observable information and pragmatic output respectively (which can be the KL divergence, cross entropy, the mean square error (MSE), etc.), $\alpha$ is a hyper-parameter to scale the value of distortion $\mathcal{D}_{\rm ob}$ for the alignment with the dynamic range of distortion $\mathcal{D}_{\rm pr}$, when their definitions are different, and $\lambda$ is a hyper-parameter to tradeoff the observation information and
pragmatic information.
\end{definition}

We would like to emphasize that semantic information is determined by specific tasks, instead of being an intrinsic feature of the source. Therefore, we consider both the observable information and pragmatic information simultaneously in the definition of the loss function. In addition, our defined SD is a generic metric where $\mathcal{D}_{\rm pr}$ can be extended to multi-task or multi-semantic scenarios.

With the given training dataset, we can define the empirical form of the loss function as follows. 
\begin{definition}
Given a training batch $\mathcal{T}$ as a collection of training sample $T=(K,Z,\widehat{K},\widehat{Z})$, the \textit{empirical semantic distortion (ESD)} is defined as
\begin{align}
\mathcal{L}_{\{\boldsymbol{\theta}_1,\boldsymbol{\theta}_2\},\mathbb{E}\sim \mathcal{T}} \triangleq & \lambda \alpha \mathbb{E}_{\sim \mathcal{T}}\left[\mathcal{D}_{\rm ob}(K,\widehat{K})\right] \nonumber \\
& + (1 -\lambda) \mathbb{E}_{\sim \mathcal{T}}\left[\mathcal{D}_{\rm pr}(Z,\widehat{Z})\right],
\label{eq:ESKL}
\end{align}
where ${\mathbb{E}_{\sim \mathcal{T}}}(\cdot)$ is the expectation w.r.t. the empirical distribution of the training batch $\mathcal{T}$. 
\end{definition}

Another important and practical issue is related to the hyper-parameter $\lambda$ in \eqref{eq:coding}. The hyper-parameter 
$\lambda$ balances the pragmatic information and the observation information. 
The main difficulty is that the tradeoff ratio is unclear. Thus, $\lambda$ can only be adjusted case by case manually. This is similar to the way hyper-parameters are normally treated in machine learning. 
In this paper, we propose a bouncy way to define the initial value of $\lambda$.

\begin{definition}
\label{def3}
A bouncy tradeoff hyper-parameter $\lambda_s$ is defined as
\begin{equation}
    \lambda_s=1-\textit{CR}=1-\frac{\log |\Gamma_{\mathcal{X}}|}{\log |\Gamma_{\mathcal{K}}|},
\end{equation}
where $\Gamma_{\mathcal{K}}$ and $\Gamma_{\mathcal{X}}$ stand for the alphabet of observable $K$ and the encoded data $X$.
\end{definition}

The $\lambda_s$ gives us an initial value of hyper-parameter to start adjusting from, but not a fixed value in practical use. The reason to pick this value is as follows. 
As a lossy joint source-channel coding with a limited coding rate, the compression rate of semantic communication takes values from $[0,1]$. Besides, when the compression rate is large, our semantic encoder has the option to encode the semantic information, as well as the non-semantic information, and its performance converges to the traditional encoder. While the compression rate is small, the semantic encoder can give a high priority to the semantic information, and converge to a pragmatic encoder correspondingly. \par

According to the above definitions, the joint optimization problem of $f_{\boldsymbol{\theta}_1}$ and $g_{\boldsymbol{\theta}_2}$ for a given compression rate constraint $\textit{CR}_0$ can be written as
\begin{align}
\min_{\boldsymbol{\theta}_1,\boldsymbol{\theta}_2} \quad & \lambda \alpha \mathbb{E}_{\sim \mathcal{T}}\!\left[\mathcal{D}_{\rm ob}(K,\widehat{K})\right]
\!+\! (1\! -\!\lambda) \mathbb{E}_{\sim \mathcal{T}}\!\left[\mathcal{D}_{\rm pr}(Z,\widehat{Z})\right], \label{eq:opt} \\
\text{s.t.} \quad & \textit{CR} = \textit{CR}_0. \nonumber
\end{align} \par

The specific function of distortion measure $\mathcal{D}_{\rm{pr}}$ varies with pragmatic tasks.
In previous works, pragmatic tasks can be classification \cite{krizhevsky2012imagenet}, detection \cite{zhao2019object}, segmentation \cite{long2015fully}, generation \cite{brock2018large}, etc. Generally, the pragmatic tasks can be sorted into two categories, based on whether the output of the pragmatic function can be regarded as a discrete random variable or not.   \par

When the pragmatic output is a discrete random variable wit a finite alphabet, for example, the image classification task, $\mathcal{D}_{\rm{ob}}$ is often chosen to be MSE and $\mathcal{D}_{\rm{pr}}$ is often chosen to be the cross entropy (CE). Hence the optimization goal in \eqref{eq:opt} can then be re-written as
\begin{align}
\min_{\boldsymbol{\theta}_1,\boldsymbol{\theta}_2} \quad & \lambda \alpha {\underbrace{\mathbb{E}_{\sim \mathcal{T}}\left[\mathcal{D}_{\textit{MSE}}(K,g_{\boldsymbol{\theta}_2}\left(f_{\boldsymbol{\theta}_1}\left(K\right)\right))\right]}_{\text{recovery task for observable information}}} \nonumber \\
& + (1 - \lambda) {\underbrace{\mathbb{E}_{\sim \mathcal{T}}\left[\mathcal{D}_{\textit{CE}}(Z,\phi\left(g_{\boldsymbol{\theta}_2}\left(f_{\boldsymbol{\theta}_1}\left(K\right)\right)\right))\right]}_{\text{classification task for pragmatic information}}}, \label{eq:optm_sce_nn_ini}
\end{align}
where $ \mathcal{D}_{\textit{MSE}} $ and $ \mathcal{D}_{\textit{CE}} $ are MSE and CE w.r.t. the empirical distribution of the training batch $ \mathcal{T} $, respectively. By taking the specific expressions of $ \mathcal{D}_{\textit{MSE}} $ and $ \mathcal{D}_{\textit{CE}} $ into \eqref{eq:optm_sce_nn_ini}, ESD for image classification task can be represented as
\begin{align}
\mathcal{L}'_{\{\boldsymbol{\theta}_1,\boldsymbol{\theta}_2\},\mathbb{E}\sim \mathcal{T}} = & \sum_{T \in \mathcal{T}}  \lambda \alpha \Vert K - \widehat{K} \Vert_2 \nonumber \\
& - (1 - \lambda) \sum_i p(Z^i) \log q(Z^i),
\label{eq:optm_sce_nn_after}
\end{align}
where $p(Z^i)$ and $q(Z^i)$ are the empirical likelihood probability that $K$ and $\widehat{K}$ are classified to the $i$-th category respectively. 

While the pragmatic output is not a discrete random variable with a finite alphabet, for example, as in image segmentation tasks, we set both $ \mathcal{D}_{\rm ob} $ and $ \mathcal{D}_{\rm pr} $ to be MSE. Hence the ESD can be written as
\begin{equation}
\mathcal{L}''_{\{\boldsymbol{\theta}_1,\!\boldsymbol{\theta}_2\},\mathbb{E}\sim \mathcal{T}}\!= \!\sum_{T \in \mathcal{T}} \!\left[\lambda \alpha \Vert K \!-\! \widehat{K} \Vert_2 \!-\! (1 \!-\! \lambda) \Vert Z \!-\! \widehat{Z} \Vert_2\right]\!,\label{continuous-loss}
\end{equation}
where $Z$ and $\widehat{Z}$ are matrices that are used to distinguish different categories and have the same shape as $K$ and $\widehat{K}$. Numerical examples with image segmentation tasks can be found in Section \ref{sec:ex_pre} as well, which uses \eqref{continuous-loss} as the loss function.

The communication cost during the training process is related to the size of raw data, compression rate, training epochs, and the complexity of the decoder network. In order to reduce the communication overhead during the training process, the semantic encoder and decoder networks can be pre-trained locally by minimizing the observation reconstruction loss $\mathbb{E}_{\sim \mathcal{T}}\!\left[\mathcal{D}_{\rm ob}(K,\widehat{K})\right]$ at the receiver. When the pre-training is completed, the receiver sends the parameters of the encoder network to the transmitter. Then, the transmitter and the receiver train the networks collectively with the whole loss $\mathcal{L}_{\{\boldsymbol{\theta}_1,\boldsymbol{\theta}_2\},\mathbb{E}\sim \mathcal{T}}$.

\section{The Data Adaptation Network}
\label{sec:DA}
During the working stage of the considered semantic communication system, when the observed dataset $\cal{S}$ is different from the library dataset $\cal{K}$, we propose data adaptation to transfer the observable data to the library data without re-training the semantic coding networks using the domain adaptation (DA) technique.
In this section, we first introduce the architecture and algorithm of the data adaptation network and then analyze its performance and feasibility.

\subsection{Domain Adaptation Architecture}
\label{sec:arch_DA}
In this subsection, we apply cycle GAN (CGAN) \cite{zhu2017unpaired} based domain adaptation in transfer learning to realize the function of data adaptation. It includes a novel architecture using the corresponding loss function design. The main purpose of this NN is to convert the observable data into a similar form of library data so that the well-trained semantic coding network can be re-used without further training.

In \cite{xie2020deep}, transfer learning is also applied to overcome the instability of the semantic communication system. Their strategy is to freeze some parameters and only re-train the rest part of them, so the re-training cost in communication and computation is reduced. 
Compared to the previous work, there are two major differences in our method. First, we introduce the domain adaptation NN to pre-process the actually observed data, while keeping the whole semantic coding network unchanged. This domain adaptation NN can be trained locally at the transmitter end, without any communication need with the receiver. Second, the CGAN architecture design of our method is for the purpose of image transmission, which has proved its performance advantages in image processing \cite{zhu2017unpaired,isola2017image,yoo2016pixel}. On the other hand, the transfer learning method proposed in \cite{xie2020deep} is designed based on the architecture of the transformer network, which is used for NLP tasks.

\begin{figure*}[t]
\centering
\includegraphics[scale=0.25]{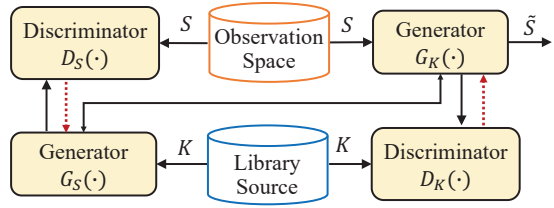}
\caption{Illustration of the data adaptation part in the proposed semantic communication system.}
\vspace{-10pt}
\label{fig:system_adaption}
\end{figure*}

The architecture of the data adaptation network is shown in Fig. \ref{fig:system_adaption}. 
There is a function converting the observed dataset into a similar representation of the library dataset called generator $G_{K}(\cdot)$. There is also a discriminator function $D_K(\cdot)$ to distinguish the library data $K$ and the data converted from observable data $G_K(S)$. When the discriminator is unable to tell these two kinds of data apart, then the network is trained well. Besides, there is also a reconverting function called generator $G_{S}(\cdot)$ and the corresponding discriminator $D_S(\cdot)$. The reconverting function is good enough if it can successfully fool the discriminator $D_S(\cdot)$ to confuse the observable data and the data reconverted from library data. The two discriminators feedback the differences between the generated data and the ground-truth data to guide the updates of the two generators, as shown in the red line. Note that though two pairs of generators and discriminators are utilized in the training process of DA, only $G_K(S)$ is needed in the inference stage.

We can see here in our architecture, that the generator function and the discriminator function play the role of two adversaries. Over the competition of the generator function and the discriminator function, the output of the converter becomes more and more similar to the library data, and the output of the generator becomes more and more similar to the observable data. Therefore, the GAN is applied as the data adaptation network in CGAN. 
Correspondingly, our data adaptation network can have the following advantages with the deployment of CGAN.

\begin{itemize}
    \item \textit{Does not require too many labeled training samples.} In the scenario where the transmitter does not have a sufficient number of pragmatic outputs given by the receiver, this approach allows the transmitter to train GANs. For example, self-supervised learning methods for CGAN do not need labeled data, while semi-supervised learning for SGAN only needs a small amount of labeled data.
    \item \textit{Can re-use the semantic encoders}. Since the encoder and decoder for the previous dataset already exist, existing coders can continue to be applied if the new dataset is converted to the previous dataset. This method saves a lot of communication costs in online learning for the new semantic coders and the pragmatic function. 
    \item \textit{Make semantic communication more scalable.} Poor scalability is a common issue in many existing semantic communication systems. For datasets with similar semantic domains, the same framework and different individual modules can be used to handle the tasks for these datasets.
\end{itemize}

Now, we introduce the detailed loss function design of the data adaptation network. In this paper, we define the library data $K\in {\cal{K}} \sim p_{\rm{lib}}(k)$ in the source domain and the observable data $S\in {\cal{S}}\sim p_{\rm{pro}}(s)$ in the target domain.
The goal of CGAN is to learn two functions $G_K(\cdot)$ and $G_{S}(\cdot)$ to implement two mappings, i.e., $G_{S}:\Gamma_{\mathcal{K}}\rightarrow \Gamma_{\mathcal{S}}$ and $G_K:\Gamma_{\mathcal{S}}\rightarrow \Gamma_{\mathcal{K}}$. To accomplish this target, two adversarial discriminators $D_K$ and $D_{S}$ are trained at the same time. The discriminator $D_K$ aims to distinguish the real data $K$ and the corresponding generated data $G_K(S)$, and $D_{S}$ works correspondingly.  \par

In conventional GAN \cite{goodfellow2014generative}, the adversarial loss is often utilized to optimize the generator and the discriminator simultaneously. Specifically, the adversarial loss between $G_S$ and $D_S$, which is denoted as $ \mathcal{L}_{S,\rm GAN}(G_S,D_S,K,S) $, is defined in \eqref{eq:loss_gan}. Meanwhile, the adversarial loss between $G_K$ and $D_K$, which is denoted as $ \mathcal{L}_{K,\rm GAN}(G_K,D_K,K,S) $, is defined accordingly in \eqref{eq:loss_gan_2}. \par

\begin{figure*}[!t]
\begin{align}
&\mathcal{L}_{S,\rm GAN}(G_S,D_S,K,S) = \mathbb{E}_{S \sim p_{\rm pro}(s)}[\log D_S(S)] + \mathbb{E}_{K\sim p_{\rm lib}(k)}\left[\log \left(1 -D_S(G_S(K))\right)\right] \\
&{\kern 97pt} =  \sum_{\mathcal{S}} p_{\rm pro}(s)\log D_S(S) + \sum_{\mathcal{K}} p_{\rm lib}(k)\log D_S(G_S(K)) \label{eq:loss_gan} \\
&\mathcal{L}_{K,\rm GAN}(G_K,D_K,K,S)  = \sum_{\mathcal{K}} p_{\rm lib}(k)\log D_K(K) + \sum_{\mathcal{S}} p_{\rm pro}(s)\log D_K(G_K(S)) \label{eq:loss_gan_2} \\
& \mathcal{L}_{\rm CGAN}(G_K, D_K, G_{S}, D_{S}, K, S) = \mathcal{L}_{S,\rm GAN}(G_K,D_K,K,S) + \mathcal{L}_{K,\rm GAN}(G_{S},D_{S},K,S) + \mathcal{L}_{\rm cycle}(G_K,G_{S},K,S) \label{eq:loss_cgan}
\end{align}
\hrule
\end{figure*}

The two forward cycles in CGAN are $K\rightarrow G_{S}(K)\rightarrow G_K\left(G_{S}(K)\right)\approx K$ and $S\rightarrow G_K(S)\rightarrow G_{S}\left(G_K(S)\right)\approx S$. The loss function of the forward cycles in CGAN is set to be
\begin{align}
    \mathcal{L}_{\rm cycle}(G_K, G_{S}, & K, S)  \!= \!\mathbb{E}_{K\sim p_{\rm lib}(k)} \left[\Vert G_K\left(G_{S}(K)\right) \!-\! K \Vert_1\right] \nonumber \\
    & + \mathbb{E}_{S\sim p_{\rm \rm pro}(s)} \left[\Vert G_{S}\left(G_K(S)\right) \!-\! S \Vert_1\right],
\label{eq:loss_cycle}
\end{align}
where $\Vert\cdot\Vert_1$ represents $L_1$-norm. The whole objective function can then be written as in  \eqref{eq:loss_cgan}, where $ D_K $ and $ D_{S} $ aim to maximize $ \mathcal{L}_{\rm CGAN}(G_K, D_K, G_{S}, D_{S}, K, S) $, while $ G_K $ and $ G_{S} $ need to minimize it. That is,
\begin{equation}
    \min_{G_{S}} \max_{D_{S}} \min_{G_K} \max_{D_K} \mathcal{L}_{\rm CGAN}(G_K, D_K, G_{S}, D_{S}, K, S).
    \label{eq:object_cgan}
\end{equation} \par

The optimization problem in  \eqref{eq:object_cgan} can be iteratively solved until the training stop condition is met. We present the training process of CGAN in Algorithm \ref{DA_traning}. \par

As a remark, semi-supervised GAN (SGAN) \cite{odena2016semi} can also be used here. CGAN can be used for cases when there is no label with the target data, while SGAN is suitable for cases when there are a few labels. There are also several other differences between the deployment of SGAN and CGAN, but in general, these two networks are quite similar. Hence, in this paper, the detailed algorithm of SGAN is omitted.\par

\begin{algorithm}[t]
% \fontsize{9pt}{11pt}\selectfont
\fontsize{10pt}{11pt}\selectfont
\caption{Training algorithm for the data adaptation networks.}
\label{DA_traning}
\begin{algorithmic}[1]
\STATE Initialize the two generators ${G_K,G_{S}}$, two discriminators ${D_K,D_{S}}$, epoch $t=1$, and batch size $ V $.

\WHILE{the training stop condition is not met}
\STATE Randomly select $V$ samples $K$ in ${\cal{K}}\sim p_{\rm{lib}}(k)$ and $V$ samples $S$ in ${\cal{S}}\sim p_{\rm{pro}}(s)$.

\STATE Generate fake pairs $\left[G_{S}(K),G_K\left(G_{S}(K)\right)\right]$,  $\left[G_K(S),G_{S}\left(G_K(S)\right)\right]$.

\STATE Utilize $D_{S}$ and $D_{K}$ to get $\mathcal{L}_{S,\rm GAN}$ and $\mathcal{L}_{K,\rm GAN}$, respectively.

\STATE Calculate $\mathcal{L}_{\rm cycle}$ by  \eqref{eq:object_cgan}.

\STATE Combining $\mathcal{L}_{S,\rm GAN}$, $\mathcal{L}_{K,\rm GAN}$, and $\mathcal{L}_{\rm cycle}$ to get $\mathcal{L}_{\rm CGAN}$.
\STATE Update $\{G_K,G_{S}\}$ by some iterative optimization algorithm to minimize $\mathcal{L}_{\rm CGAN}$.

\STATE Update $\{D_K,D_{S}\}$ by some iterative optimization algorithm to maximize $\mathcal{L}_{\rm CGAN}$.

\ENDWHILE

\end{algorithmic}
\end{algorithm}

\subsection{Similarity Measure Between Two Datasets}
\label{sec:analysis}
This subsection first provides a measure function of the similarity between the library dataset and the observed dataset for some particular pragmatic task. This measure function can be an indicator of the potential gain of using DA. In general, the less similar the two datasets are, the higher the potential gain can DA bring. Then we also introduce a machine learning method to approximate this measure function for numerical evaluation.

We define the domain as a pair consisting of a distribution $p(\cdot)$ on inputs and a labeling function $\upsilon(\cdot)$ mapping the inputs to $[0,1]^n$, where $n$ is the feature dimensions. For example, when the pragmatic task is digit classification, the function $\upsilon(\cdot)$ can be the ground truth of whether the image is of a particular number. The domain of the library dataset and observation space can be denoted as $\{p_{\rm lib},\upsilon_{\rm lib}\}$ and $\{p_{\rm pro},\upsilon_{\rm pro}\}$, respectively. We also define the \textit{hypothesis} as the indicator function $h: \mathcal{K}\rightarrow \{0,1\}^n$, which can be regarded as an extreme case of the labeling function. Then the difference between the two labeling functions $\upsilon$ and $\upsilon'$ over the library set can be defined as \cite{ben2010theory}
\vspace{-7pt}
\begin{equation}
    \epsilon_{\rm lib}(\upsilon,\upsilon')\triangleq \mathbb{E}_{k\sim p_{\rm lib}}\left[|\upsilon(k)-\upsilon'(k)|\right],
    \vspace{-7pt}
\end{equation}
and $\epsilon_{\rm pro}(\upsilon,\upsilon')$ can be defined accordingly. Moreover, a metric of the difference between two datasets called \textit{$\mathcal{H}\Delta\mathcal{H}$-divergence} can also be defined as follow.

\begin{definition} \label{def4}
The discrepancy measure between the domains of the library and observed datasets, which is called \textit{$\mathcal{H}\Delta\mathcal{H}$-divergence}, can be defined as
\begin{align}
d_{\mathcal{H}\Delta\mathcal{H}} & \triangleq 2 \sup_{h,h'\in \mathcal{H}}\big| \mathbb{E}_{k\sim p_{\rm lib}}\mathbf{1}[h(k)\neq h'(k)] \nonumber \\
& \quad\quad \quad -\mathbb{E}_{s\sim p_{\rm pro}}\mathbf{1}[h(s)\neq h'(s)] \big| \\
& = 2\sup_{h,h'\in \mathcal{H}}|\epsilon_{k\sim p_{\rm lib}}(h,h')-\epsilon_{s\sim p_{\rm pro}}(h,h')| \\
& \geq 2|\epsilon_{k\sim p_{\rm lib}}(h,h')-\epsilon_{s\sim p_{\rm pro}}(h,h')|,
\label{eq:discre}
\end{align}
where $\mathbf{1}[\cdot]$ is the indicator function, $h$ and $h'$ are two hypotheses, $\mathcal{H}$ is the hypothesis space. Besides, the \textit{symmetric difference hypothesis space} $\mathcal{H}\Delta\mathcal{H}$ is the set of hypotheses that satisfy
\begin{equation}
    \beta \in \mathcal{H}\Delta\mathcal{H} \Leftrightarrow \beta(k) = h(k) \oplus h'(k), 
\end{equation}
for all $h,h'\in \mathcal{H}$, where $\oplus$ is the XOR function and $\beta$ is an indicator function in the symmetric difference hypothesis space $\mathcal{H}\Delta\mathcal{H}$.
\end{definition}

Here the lower bound of \textit{$\mathcal{H}\Delta\mathcal{H}$-divergence} in  \eqref{eq:discre} can be regarded as the total variation between two probability functions, which denote the error probability of decoding the pragmatic output of library data and observed data respectively. Hence,  \eqref{eq:discre} means that for a fixed hypothesis space $\mathcal{H}$, $d_{\mathcal{H}\Delta\mathcal{H}}(\mathcal{K},\mathcal{S})$ is the intrinsic difference between the domains of the library and observed datasets, which is fixed and determined by the characteristics of the data distributions.  \par

Specifically, if the pragmatic outputs in $\mathcal{K}$ and $\mathcal{S}$ are almost the same, $d_{\mathcal{H}\Delta\mathcal{H}}(\mathcal{K},\mathcal{S})$ will be small. Otherwise, this item will increase with the difference in the pragmatic outputs in the two domains. Therefore, we select datasets with the same category as each other in the experiment, such as the USPS and MNIST datasets. Since the data in both domains are available at the transmitter, the transmitter can easily figure out the differences between different domains. In general, \eqref{eq:discre} provides guidance on when to trigger the data adaptation network and whether the semantic communication system needs to be retrained. \par

Besides the lower bound introduced above, $\mathcal{H}\Delta\mathcal{H}$-divergence between $p_{\rm lib}$ and $p_{\rm pro}$ can also be tightly approximated by the so-called proxy $\mathcal{A}$-distance (PAD). PAD follows a machine learning method. Specifically, we first construct a new dataset as
\begin{equation}
    \widetilde{K}=\{(k_i,0)\}_{k_i\sim p_{\rm lib}}\cup\{(s_j,1)\}_{s_j\sim p_{\rm \rm pro}},
\end{equation}
where $\{(k_i,0)\}_{k_i \sim p_{\rm lib}}$ denotes the set of samples following distribution $p_{\rm lib}$, which are labeled as zero, and $\{(s_j,1)\}_{s_j \sim p_{\rm \rm pro}}$ is defined similarly. Then, a binary classifier is trained on a subset of the newly constructed dataset $\widetilde{K}$. As suggested by previous works \cite{ben2006analysis,ben2010theory}, linear classifiers can be used here, such as SVM, MLP, linear CNN, etc. Suppose the classification error is $\epsilon$, then PAD is defined as 

\begin{equation}
    d_{\mathcal{A}}\triangleq2(1-2\epsilon).
\end{equation}
It can be seen that the smaller $d_{\mathcal{A}}$ is, the more similar these two domains are. \par
Note that as a machine learning method, PAD serves as a good approximation of $\mathcal{H}\Delta\mathcal{H}$-divergence and is convenient for numerical evaluation. Some experimental results shall be given later in Section \ref{sec:ex_DA}.  \par

\section{Numerical Experiments}
\label{sec:results}
In this section, we validate the performance of the semantic coding part and the data adaptation part, respectively. 
Generally speaking, both the bit-wise performance and the semantic performance are tested under different compression rate $\textit{CR}$ at different channel conditions. 
Detailed setups will be given for each specific experiment respectively.
As a remark, according to the information bottleneck theory \cite{information_bottleneck}, compared with the convolution function of the convolutional layers or the linear function of fully connected layers, activation functions of NNs can lead to greater amounts of information loss. Therefore, the NNs of the semantic coders do not contain activation functions in our experiments. An additional normalization layer is also added at the semantic encoder output to ensure the transmit power constraint. Throughout the experiments, the signal-to-noise ratio (SNR) of the AWGN channel is set to be 3dB and 10dB, representing the low and high SNR regimes, respectively.

\subsection{Experiments on Semantic Coding Network}
\label{sec:ex_pre}
In this subsection, we evaluate our semantic coding network on three representative experiments, namely handwritten digit recognition on the MNIST dataset, image classification on the CIFAR-10 dataset, and image segmentation on the PASCAL-VOC2012 dataset, respectively. For comparison, two benchmark schemes are considered. The first one is the traditional separate source-channel coding (SSCC) which employs JPEG2000 \cite{christopoulos2000jpeg2000} for image compression and capacity-achieving codes for reliable channel transmission at the rate $\frac{1}{2}\log(1+\textit{SNR})$ bits per channel use. Hence, the compression rate of JPEG2000 should be the compression rate of our proposed neural network encoder times $\frac{1}{2}\log(1+\textit{SNR})$.
The second one is the variational auto-encoder (VAE)-based semantic communication \cite{VAE} with the KL divergence loss, which is written as the VAE-based method for convenience. Note that we choose VAE instead of a standard autoencoder since VAE can better respond to the effects of noise as shown in the existing experiments \cite{VAE,mehrasa2019variational}. \par

\subsubsection{Semantic Communication for MNIST Digit Recognition}
\label{sec:ex_MNIST}
The MNIST dataset consists of 60,000 images with each being handwritten numbers with $28 \times 28$ grayscale pixels \cite{lecun1998gradient}. 
Among the dataset, 50,000 images are used for training, and the rest 10,000 images are used for testing.

The pragmatic function $\phi(\cdot)$ used at the receiver for digit recognition is trained in advance using the ground-truth label.
In our method, the encoding and decoding neural networks are both with one fully connected layer. In the VAE-based method, the encoding and decoding neural networks are both with two fully connected layers.
We use two important criteria to evaluate the performance of different methods, namely, the accuracy of digit recognition for the pragmatic task and the peak signal-to-noise ratio (PSNR) for image reconstruction. \par

\begin{figure}[tb]
\centering
\subfigure[Test Accuracy.]{
\includegraphics[scale=0.5]{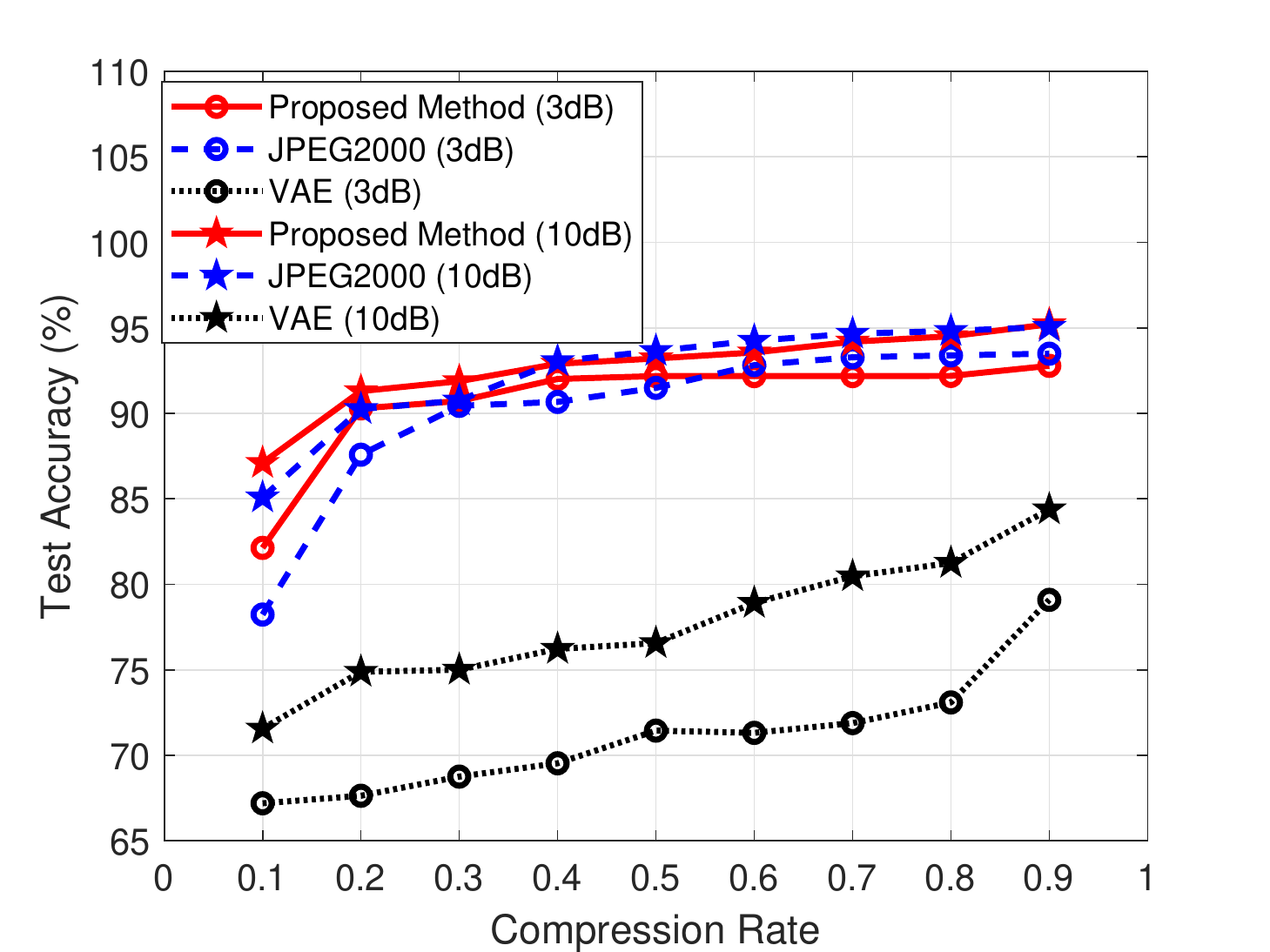}}
\subfigure[PSNR.]{
\includegraphics[scale=0.5]{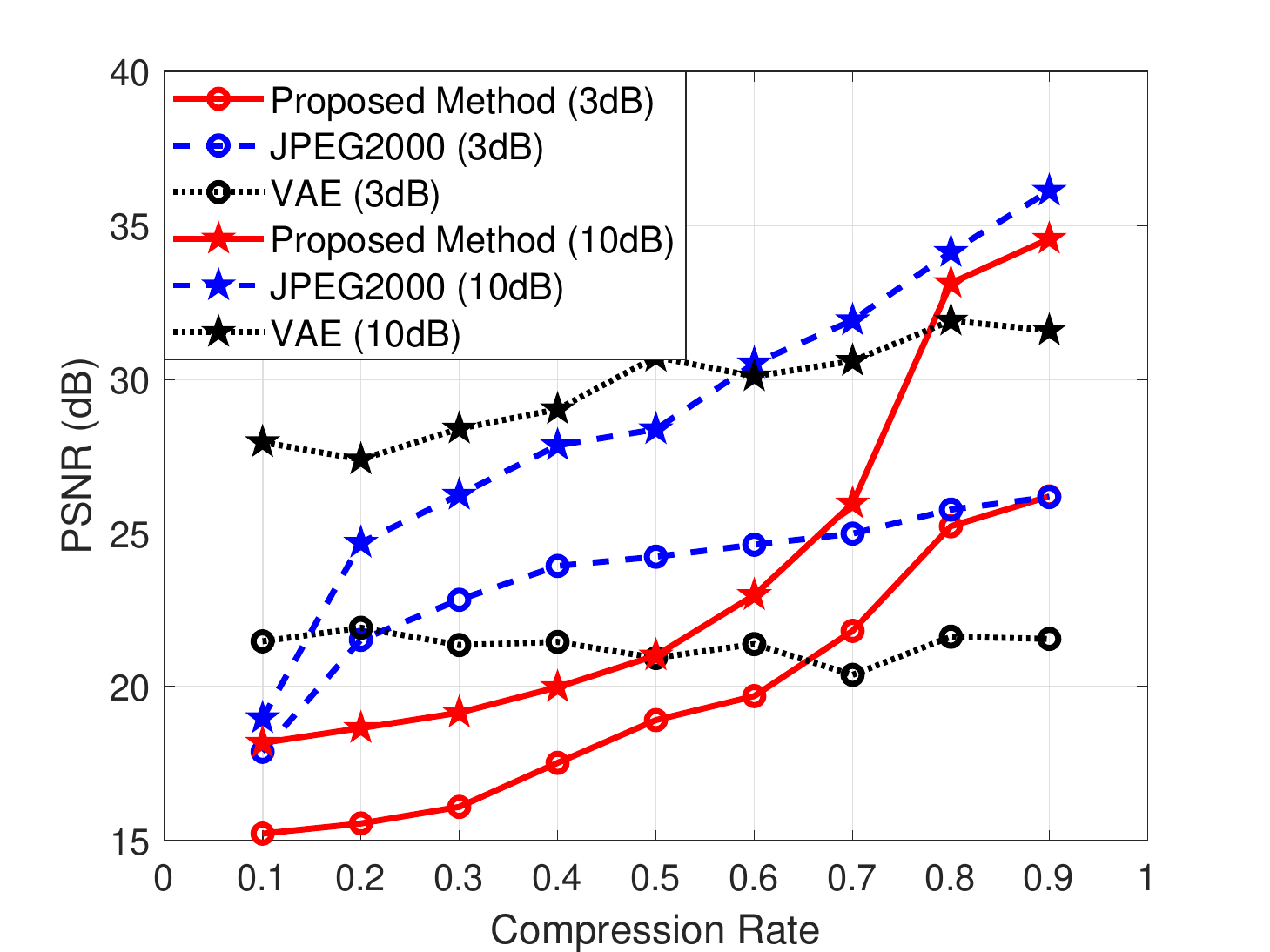}}
% \vspace{-10pt}
\caption{Performance comparison with the MNIST dataset.}
\vspace{-15pt}
\label{mnsit_diff_com}
\end{figure}

Fig. \ref{mnsit_diff_com} compares the performance of our method with benchmark schemes in terms of recognition accuracy and PSNR against compression rates. It is seen from Fig. \ref{mnsit_diff_com}(a) that, the proposed method has the highest accuracy among all the considered schemes when $\textit{CR}$ is less than 0.4, and the VAE method performs the worst at all different $\textit{CR}$. This is because the proposed loss function can weight the importance of observable information and pragmatic information according to different compression rates. When $\textit{CR}$ is low, the proposed scheme will give priority to ensuring the performance of tasks, i.e., the accuracy of digit recognition. When $\textit{CR}$ is greater than 0.2, there is a minor difference between the proposed method and the JPEG2000-based method, for both 3dB and 10dB SNR. In particular, when $\textit{CR}\in [0.1,0.4]$, the recognition accuracy of the proposed method is better, while when $\textit{CR}\in[0.6,0.9]$, the recognition accuracy of the JPEG2000-based method is slightly better. In general, the advantages of the proposed method are not obvious in this case. This is because the samples in the MNIST data set are so simple that they can be encoded at a very low compression rate with negligible information loss, making the performance advantage of JSCC very minor over SSCC.

Meanwhile, we can see from Fig. \ref{mnsit_diff_com}(b) that, the proposed method has no performance advantage in terms of PSNR at both 3dB and 10bB SNR. On the other hand, the VAE method has the best PSNR performance at 10dB SNR, and JPEG2000-based method performs the best when $\textit{CR}\in [0.2,0.9]$ at 3dB SNR and $\textit{CR}\in[0.6,0.9]$ at 10dB SNR. However, we can see the PSNR performance of our method increases with the growth of $\textit{CR}$, and gradually close to the best performed method. This phenomenon coincides with the loss function design of the proposed method. Combining with the results in Fig. \ref{mnsit_diff_com}(a), we can see the priority tradeoff between pragmatic task and image reconstruction in the proposed method. When the compression is low, the priority of pragmatic tasks will damage the performance of image reconstruction. But when $\textit{CR}$ is high, the focus of the encoder in the proposed method will return to image reconstruction, and its PSNR performance can exceed that of other methods.

\subsubsection{Semantic Communication for CIFAR-10 Image Classification} The CIFAR-10 dataset \cite{krizhevsky2009learning} consists of 60,000 RGB images with size $32\times 32$ in 10 classes, with 6,000 images per class. Among the dataset, 50,000 images are used for training and the rest 10,000 images are used for testing. The pragmatic use here is image classification. Hence, here we still use the accuracy and the PSNR as the criteria to compare. \par

\begin{figure}[tbp]
\centering
\subfigure[Test Accuracy.]{
\includegraphics[scale=0.5]{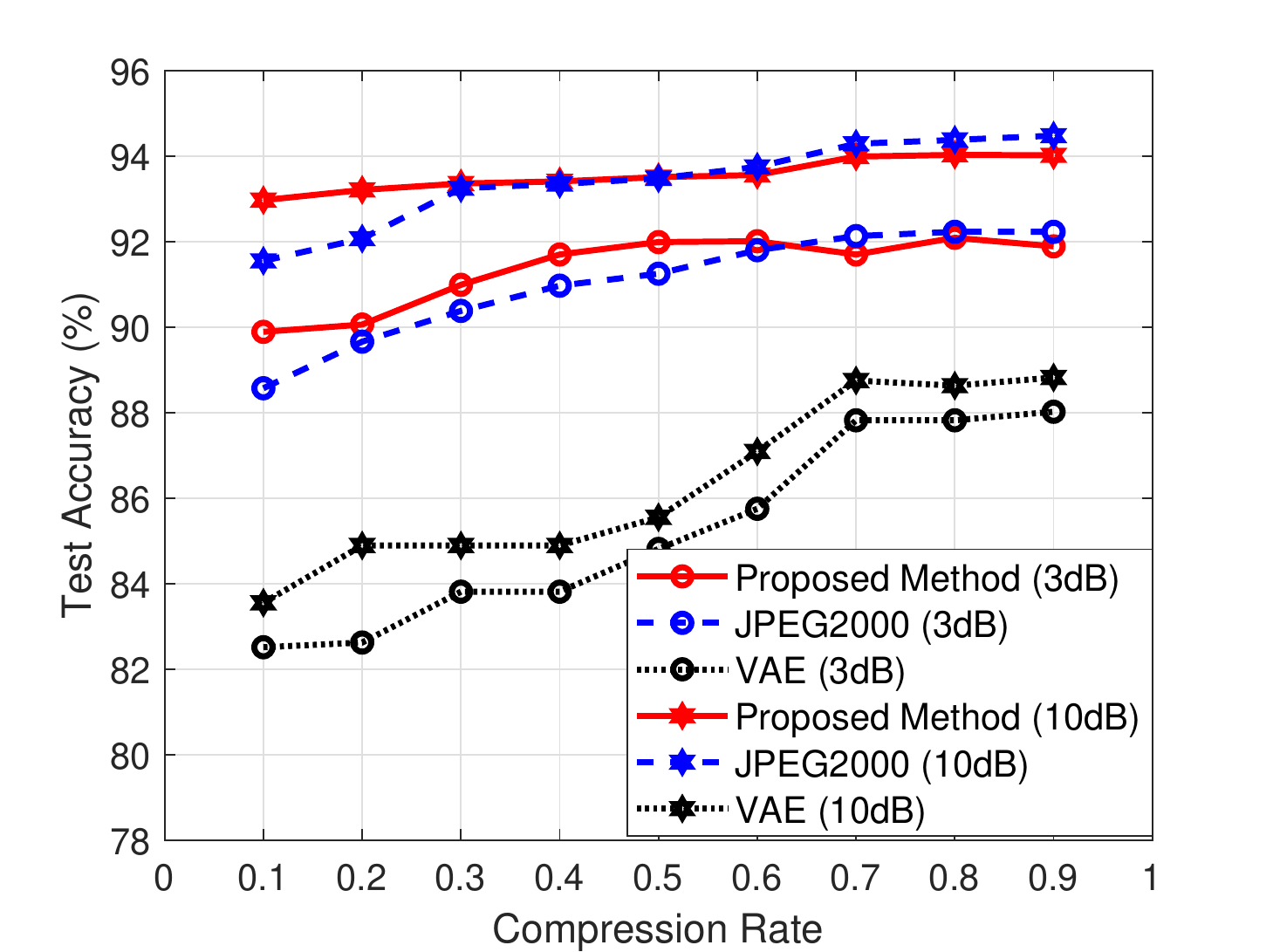}}
\subfigure[PSNR.]{
\includegraphics[scale=0.5]{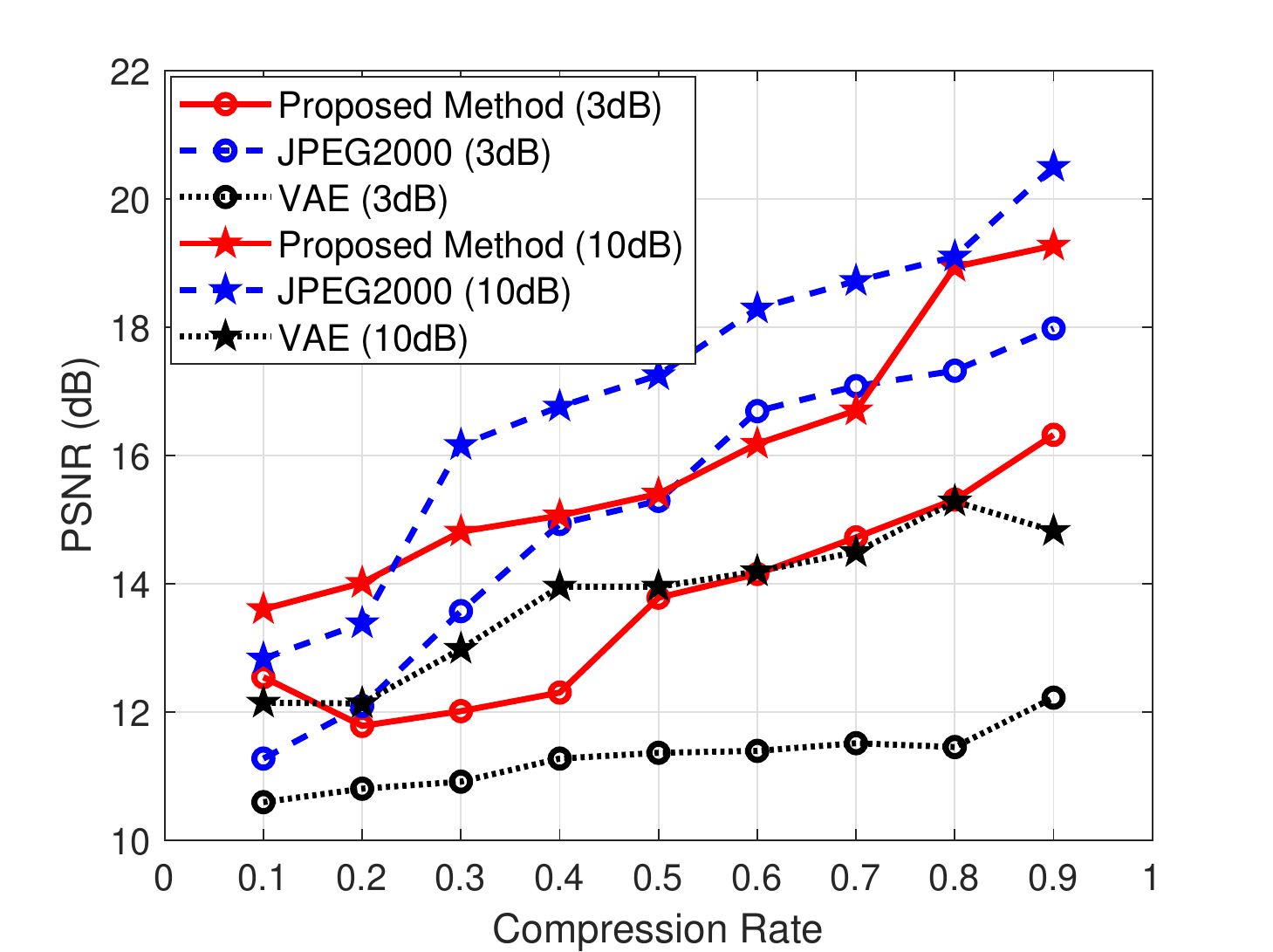}}
% \vspace{-10pt}
\caption{Performance comparison with the CIFAR10 dataset.}
\vspace{-15pt}
\label{cifar_diff_com}
\end{figure}

Fig. \ref{cifar_diff_com}(a) plots the test classification accuracy under different compression rates.
We can see that the proposed method has the best performance among the three methods with low $\textit{CR}$, and is very robust to the change of $\textit{CR}$ at both 3dB and 10dB SNR. In particular, the classification accuracy of the proposed method is still as high as $93\%$ when $\textit{CR}$ is 0.1 at 3dB SNR, while the accuracy of the JPEG2000 method and the VAE method is only $91.8\%$ and $82.4\%$ respectively, at the same compression rate and same channel condition.  
This is because the image in CIFAR10 is more complicated than that in MNIST, and classifying images in CIFAR10 is more difficult than digit recognition. Hence the classification accuracy cannot be guaranteed, unless the reconstructed image is very close to the raw image, like the situation of the JPEG2000-based method with a high compression rate, or taking image classification as a target in encoding, like the proposed method.

Fig. \ref{cifar_diff_com}(b) plots the PSNR against compression rates. It is seen that the proposed method has the best PSNR performance when $\textit{CR}$ is low. When $\textit{CR}$ is high, the JPEG2000-based method has the best PSNR performance.
We can still see the tradeoff between pragmatic performance and image reconstruction. Because the images in CIFAR10 and the coder networks used in this experiment are more complicated, it is easier to converge to a poor local optimum when modeling the distribution of these images with NNs. When $\textit{CR}$ is higher, the coder networks are more complex, which makes it harder to converge to a better local optimum. However, the advantage of the proposed method is apparent when $\textit{CR}$ is less than 0.2 at 3dB. Specifically, the tradeoff between information compression and noise reduction is more important in this case, and the proposed method adopts the end-to-end based JSCC, so it can balance them in a learning way.

\begin{figure}[tbp]
\centering
\subfigure[Raw image.]{
\includegraphics[scale=0.6]{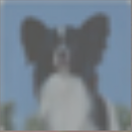}}
\subfigure[VAE, $CR=0.1$.]{
\includegraphics[scale=0.6]{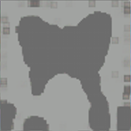}}
\subfigure[Proposed, $CR=0.1$.]{
\includegraphics[scale=0.6]{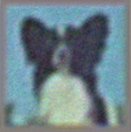}}
\subfigure[JPEG2000, $CR=0.1$.]{
\includegraphics[scale=0.41]{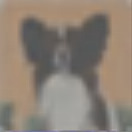}}
\subfigure[Raw image.]{
\includegraphics[scale=0.6]{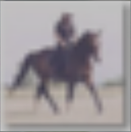}}
\subfigure[VAE, $CR=0.8$.]{
\includegraphics[scale=0.6]{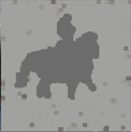}}
\subfigure[Proposed, $CR=0.8$.]{
\includegraphics[scale=0.6]{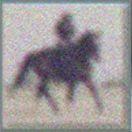}}
\subfigure[JPEG2000, $CR=0.8$.]{
\includegraphics[scale=0.41]{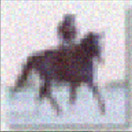}}
\caption{Raw images and images recovered by different methods with the CIFAR10 dataset.}
\vspace{-15pt}          
\label{fig:cifar1}
\end{figure}

Fig. \ref{fig:cifar1} gives visible results of the recovered images with $\textit{CR}=$0.1 and 0.8 respectively and SNR$=3$dB. It is seen that the margin of the recovered images of the proposed scheme is severely polluted, but the object in the recovered images is the clearest among all the three methods. Notice that there are many chromatic aberrations in the images recovered by the VAE-based method, while it can be seen that the proposed method and the JPEG2000-based method have similar resolutions for the recovered images. Meanwhile, the reconstructed image of the VAE method can only present approximate outlines of the objects in the raw image. These reconstructed images coincide with the intuition of three methods. The JPEG2000-based method is to transmit the contour of the image first, and then gradually transmit the data to constantly improve the image quality. So that the image is from hazy to clear display with the growth of $\textit{CR}$. Hence, the outline of its reconstructed images is visibly close to that of the raw images, while the colors subsequently filled in the reconstructed images are different from that in the raw images. The proposed method focuses on both pragmatic performance and image reconstruction, so its reconstructed images still keep the outline of raw images for the pragmatic task, but more blur than that of the JPEG2000-based method at $\textit{CR}=0.1$ due to the distraction of transmitting semantic information. These differences between the visible images of our method and JPEG2000-based method also match with the numerical results in Fig. \ref{cifar_diff_com}(b). The VAE is a semi-generative model, hence in some sense, it is more like trying to build a new image, and its reconstructed image looks the most disliked to the raw image among the three methods. \par

\begin{figure}[tb]
\centering
\subfigure[IoU.]{
\includegraphics[scale=0.5]{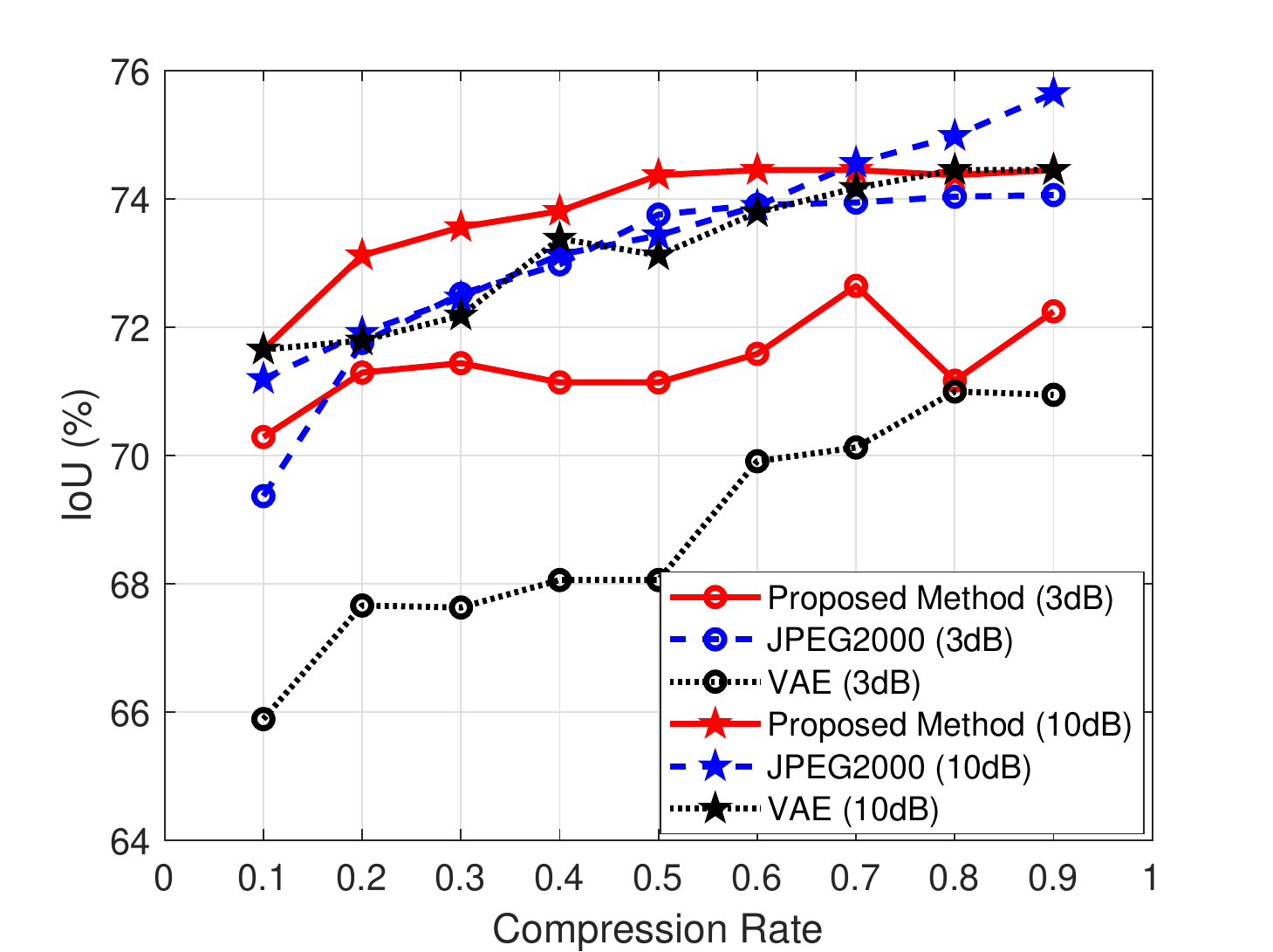}}
\subfigure[PSNR.]{
\includegraphics[scale=0.5]{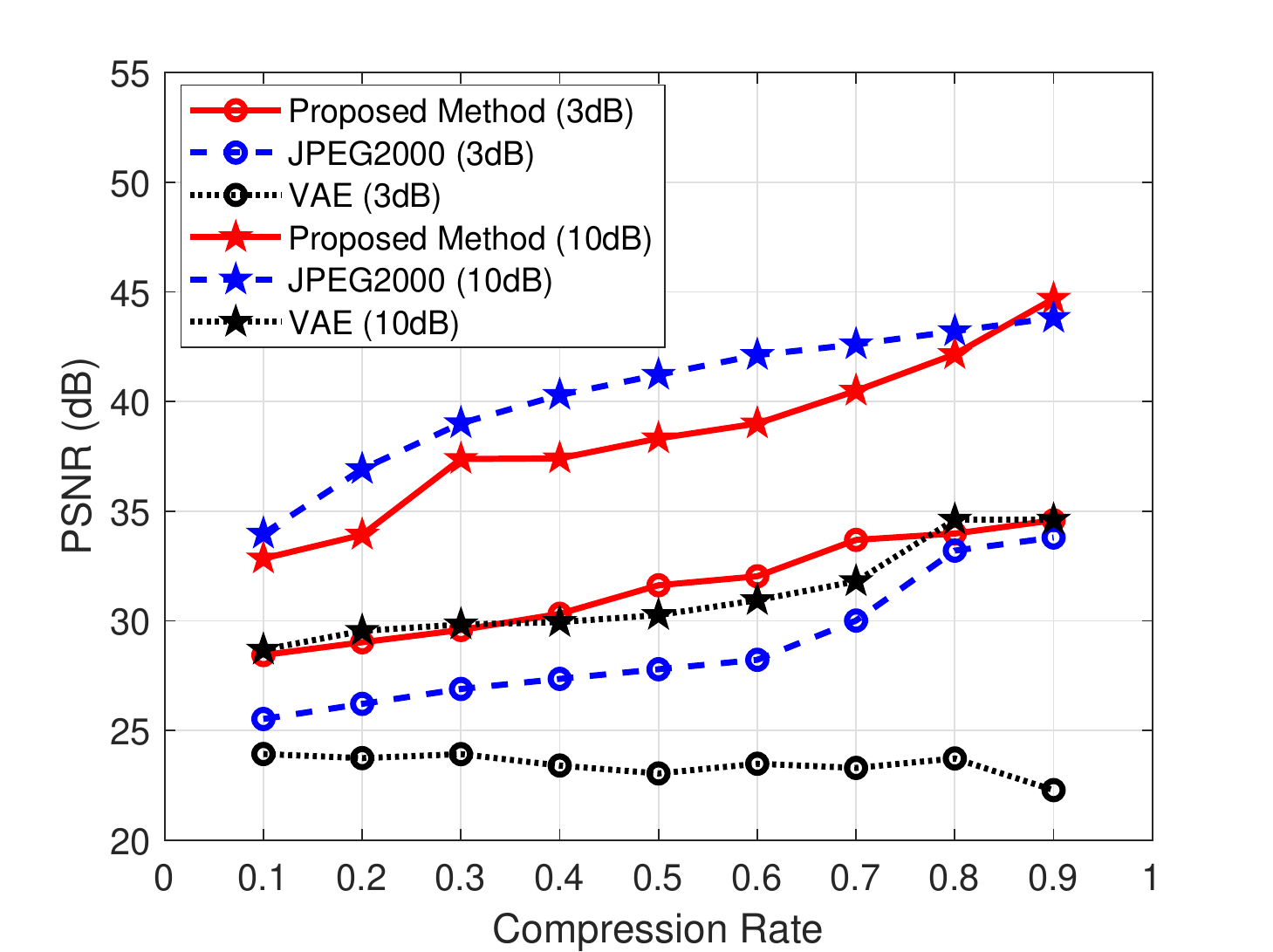}}
% \vspace{-10pt}
\caption{Performance comparison with the PSCAL-VOC2012 dataset.}
\vspace{-15pt}
\label{voc_diff_com}
\end{figure}

\begin{figure*}[b]
\centering
\subfigure[Raw image.]{
\label{fig:subfig:voc_raw0.1} 
\includegraphics[scale=0.5]{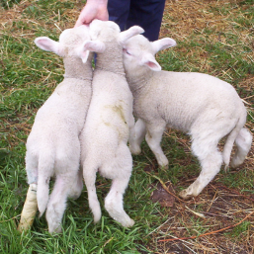}}
\hspace{0.15in}
\subfigure[VAE, $CR=0.1$.]
{
\label{fig:subfig:voc_pragmatic0.1}
\includegraphics[scale=0.5]{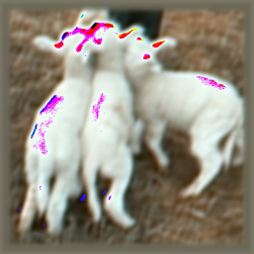}}
\hspace{0.15in}
\subfigure[Proposed, $CR=0.1$.]{
\label{fig:subfig:voc_semantic0.1}
\includegraphics[scale=0.5]{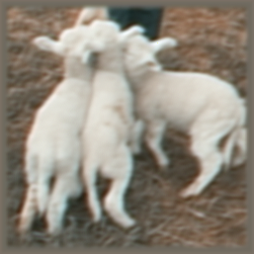}}
\hspace{0.15in}
\subfigure[JPEG2000, $CR=0.1$.]{
\label{fig:subfig:voc_data0.1}
\includegraphics[scale=0.375]{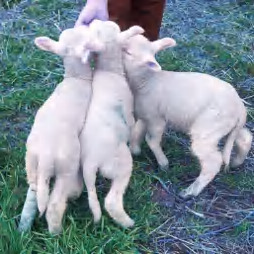}}
\subfigure[Raw image.]{
\label{fig:subfig:voc_raw} 
\includegraphics[scale=0.35]{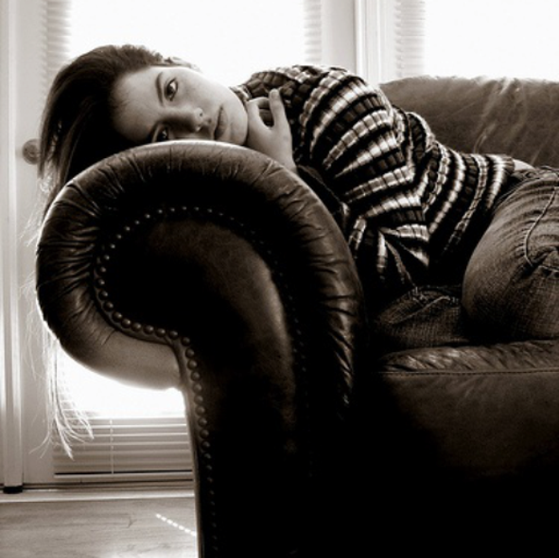}}
\hspace{0.15in}
\subfigure[VAE, $CR=0.8$.]
{
\label{fig:subfig:voc_pragmatic}
\includegraphics[scale=0.35]{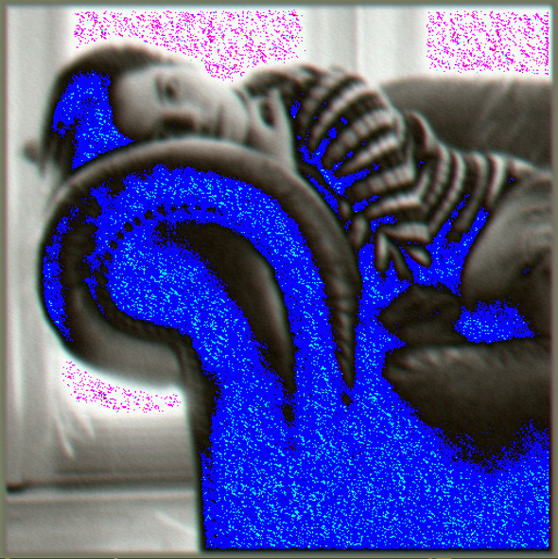}}
\hspace{0.15in}
\subfigure[Proposed, $CR=0.8$.]{
\label{fig:subfig:voc_semantic}
\includegraphics[scale=0.35]{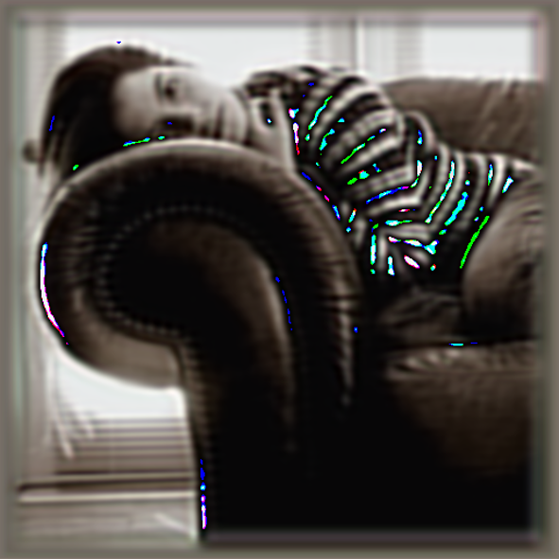}}
\hspace{0.15in}
\subfigure[JPEG2000, $CR=0.8$.]{
\label{fig:subfig:voc_data}
\includegraphics[scale=0.17]{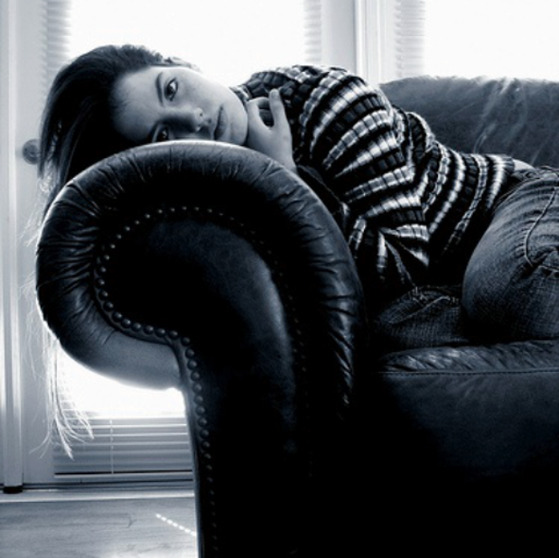}}
% \vspace{-10pt}
\caption{Raw images and images recovered by different methods with the PASCAL-VOC2012 dataset.}
% \vspace{-20pt}
\label{fig:recover_VOC} %% the pragmatic outputsfor entire figure
\end{figure*}

\subsubsection{Semantic Communication for Image Segmentation}
\label{sec:ex_NN}

The PASCAL-VOC2012 dataset consists of 2,913 RGB images with sizes $513\times 513$ in 20 categories \cite{everingham2011pascal}. Among the dataset, 10,582 images are used for training and the rest 1,449 images are used for testing.
For our method, we use a five-layer CNN and deconvolution NN as the encoder and decoder, respectively. Each layer of CNN adopts a convolution kernel with the same size of $ w \times w $, where $ w = \left\lceil \frac{\log{\lvert \Gamma_{\mathcal{X}} \rvert}}{5}\times \left(1 - \sqrt{CR}  \right) \right\rceil $. \par

Two objective criteria are used to evaluate the performance, namely, the intersection of union (IoU) and the PSNR. IoU is a widely used method to evaluate the accuracy of object detection in computer vision. It is defined as the ratio between the intersection area and the union area of the segmented parts in the recovery and raw images. Fig. \ref{voc_diff_com}(a) shows the IoU performance for all three methods under different compression rates. We can see that for such a complicated dataset and pragmatic task, our proposed methods have better pragmatic performances than the JPEG2000-based method at 10dB SNR when $\textit{CR}$ is low, and slightly worse pragmatic performances than the JPEG2000-based method at 3dB SNR. We also can see the pragmatic performance of our method has an apparent increase with the growth of $\textit{CR}$, while in the previous two experiments, the pragmatic performance is not very sensitive to the change of $\textit{CR}$. It proves that image segmentation takes far more information than the image classification task. 

However, the graph of PSNR against compression rate in Fig \ref{voc_diff_com}(b) shows that the JPEG2000-based method still has the best pixel-wise performance at 10dB SNR. The gap between the PSNR of the JPEG2000-based method and that of the proposed method becomes smaller when $\textit{CR}$ increases. However, the pixel-wise performance of the proposed method is better than that of the JPEG2000-based method at 3dB SNR and low $\textit{CR}$. This is because JSCC can outperform SSCC when the source is complex and the channel condition is poor. We can still see the tradeoff between the pragmatic task and image reconstruction in the proposed method. But since the pragmatic task occupies more information, the PSNR performance of the proposed method in this experiment is not as good as that in previous experiments.

Fig. \ref{fig:recover_VOC} gives a more direct comparison of all three methods by visible results. Each row includes the raw image and images recovered by the VAE-based method, the proposed method, and the JPEG2000-based method. The coding rate for all three encoders is fixed at 1,895,064 bits, i.e., an RGB image with a resolution of 281$\times$281. We can find that though the details of the recovered image of our proposed method are not as clear as that of the JPEG2000-based method, but it can better preserve the outline of different areas. This is another evidence that the proposed method emphasizes protecting the semantic information. As to this experiment, the outlines of different objects are the semantic information that we want to protect. There are also many chromatic aberrations in the reconstructed image of JPEG2000. Besides, the reconstructed image of JPEG2000 with $\textit{CR}$=0.1 has more contaminated patches of color than that with $\textit{CR}$=0.8.   \par
Generally, the above experiments find out that our proposed semantic coding network can have a better performance on pragmatic tasks than the JPEG2000-based method when the compression rate $\textit{CR}$ is low. However, the threshold value of the compression rate within which our semantic coding network has dominant performance varies with the complexity of the dataset and pragmatic task.

\subsection{Experiments on Data Adaptation Network}
\label{sec:ex_DA}
In this subsection, we conduct two representative experiments. In the first experiment, We use MNIST as the library dataset and use two datasets SVHN and USPS respectively as the observed dataset. 
The SVHN dataset \cite{netzer2011reading} is a real-world image dataset, which takes house numbers from Google street view images. It has thus ten categories of numbers. Besides, every sample in the SVHN dataset has $32\times 32$ three-channel pixels. On the other hand, the USPS dataset \cite{hull1994database} includes 7,291 training and 2,007 test images of handwritten digits (ten categories), and each sample has $16 \times 16$ grayscale pixels. \par

In the second experiment, CIFAR10 is used as the library dataset with STL10 being the observed dataset. 
The STL10 dataset \cite{coates2011analysis} is similar to the CIFAR10 dataset. There are fewer labeled training samples for each class in the STL10 dataset than that in the CIFAR-10 dataset. Besides, each sample in the STL10 dataset is also an RGB image and has a higher resolution ($96 \times 96$ pixels) than that in the CIFAR10 dataset. \par

Two benchmarks are used to compare with our proposed method. The first benchmark is obtained by inputting the observed data directly into the semantic coding network encoder. Since image data in different datasets have different resolutions, we use image up-sampling or down-sampling to make observed image data the same size as that in the library dataset. We call this benchmark the method without DA, whose performance relies on the scalability of the semantic coding network. Another benchmark is obtained by re-training the whole semantic coding networks with observed data, which can be regarded as an ideal upper bound of the performance, though unpractical due to the communication overhead. As a remark, all three methods are named ``dataset (No DA)", ``dataset (Retrained)" and ``dataset (DA)" in the legends of the figures, respectively. 

\subsubsection{DA-Based Semantic Communication for MNIST}

\begin{figure}[t]
\centering
\subfigure[Accuracy against epoch.]{
\label{fig:subfig:svhn_epoch}
\includegraphics[scale=0.5]{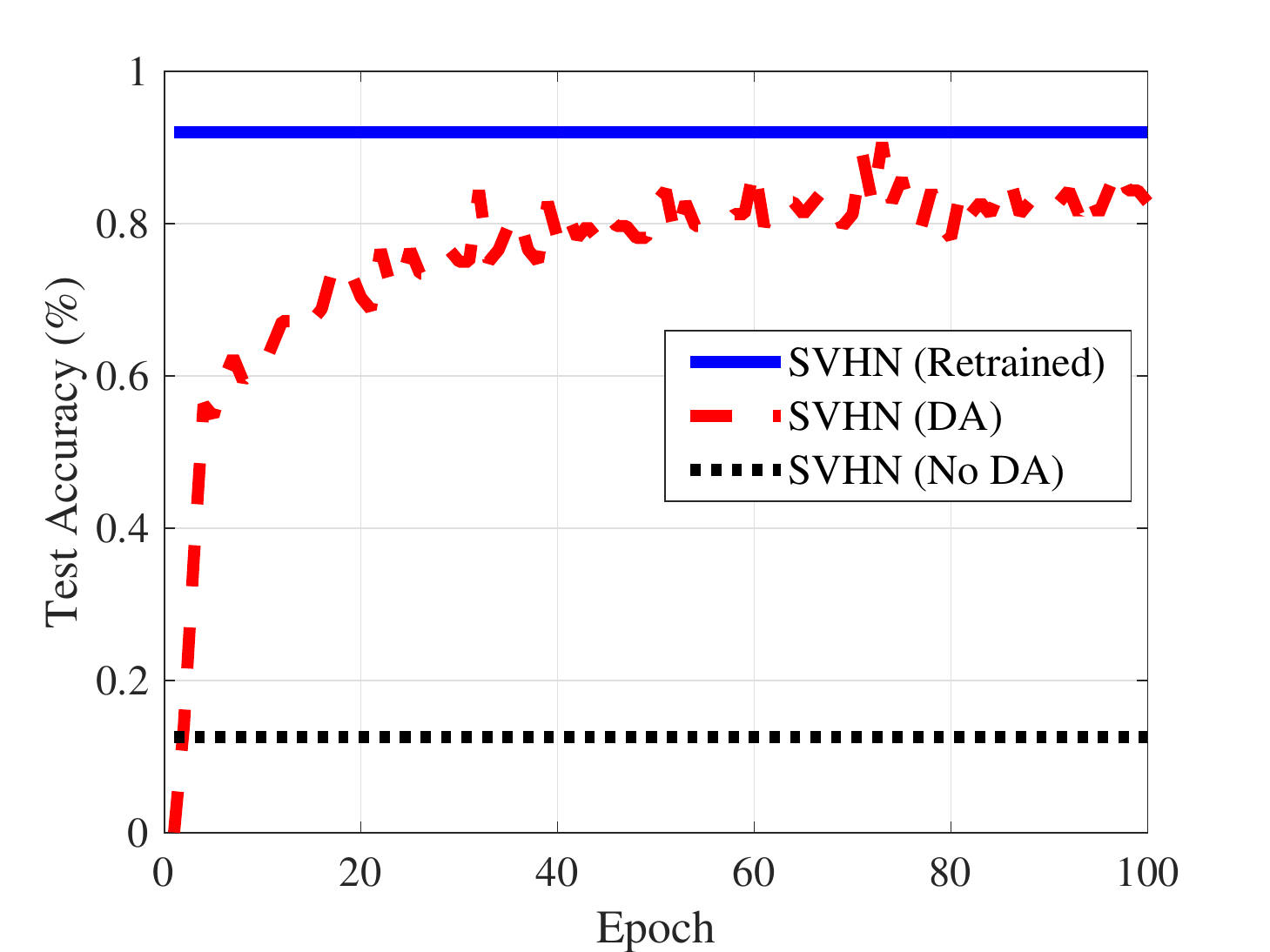}}
\subfigure[Accuracy against $CR$.]{
\label{fig:subfig:svhn_cr} %% the pragmatic outputsfor first subfigure
\includegraphics[scale=0.5]{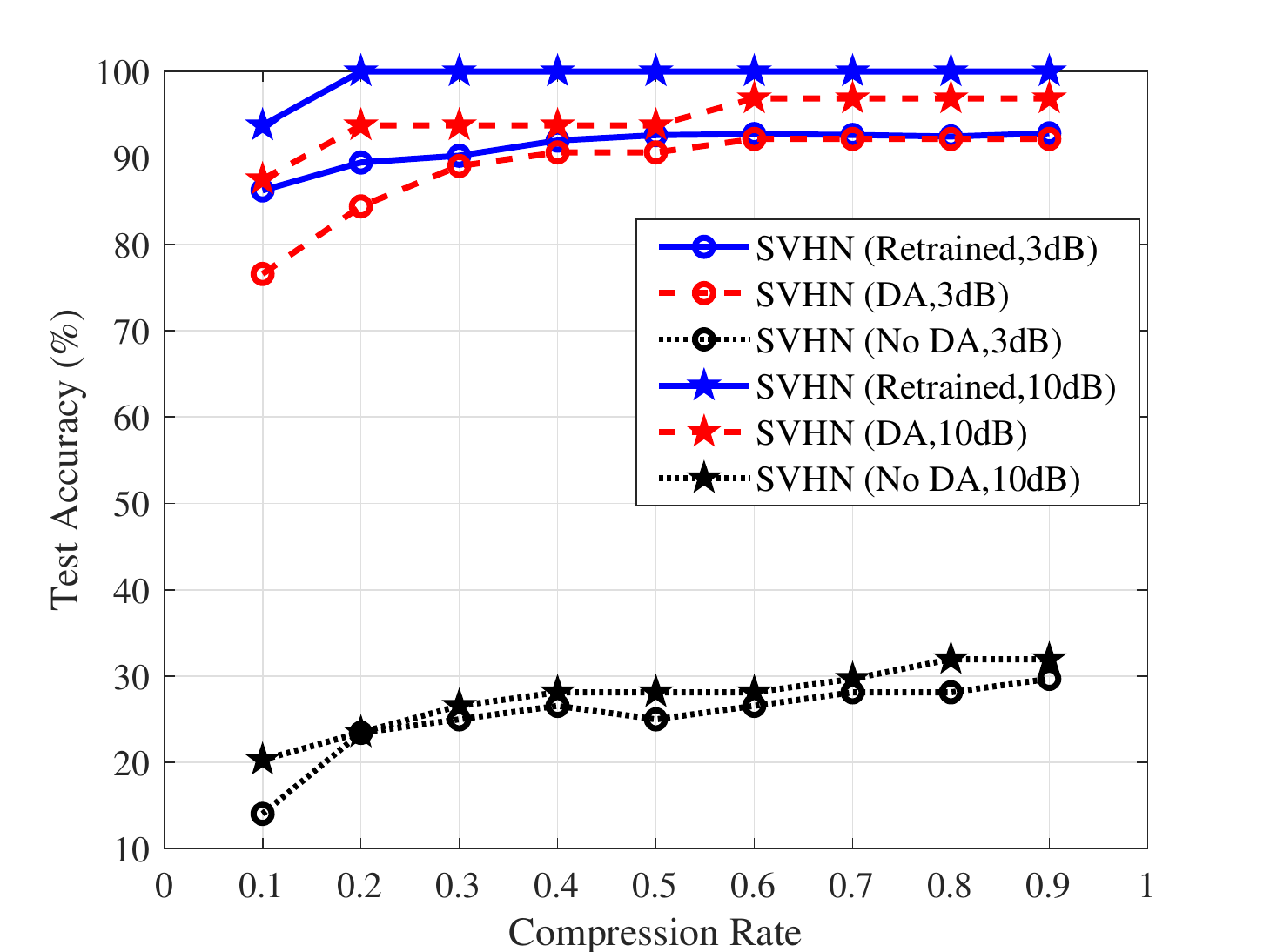}}
% \hspace{-0.3in}
\caption{The accuracy of digit recognition with DA from SVHN to MNIST.}
\vspace{-20pt}
\label{fig:transfer_mnist_SVHN}
\end{figure}

\begin{figure}[t]
\centering
\subfigure[Accuracy against epoch.]{
\label{fig:subfig:usps_epoch} %% the pragmatic outputsfor second subfigure
\includegraphics[scale=0.5]{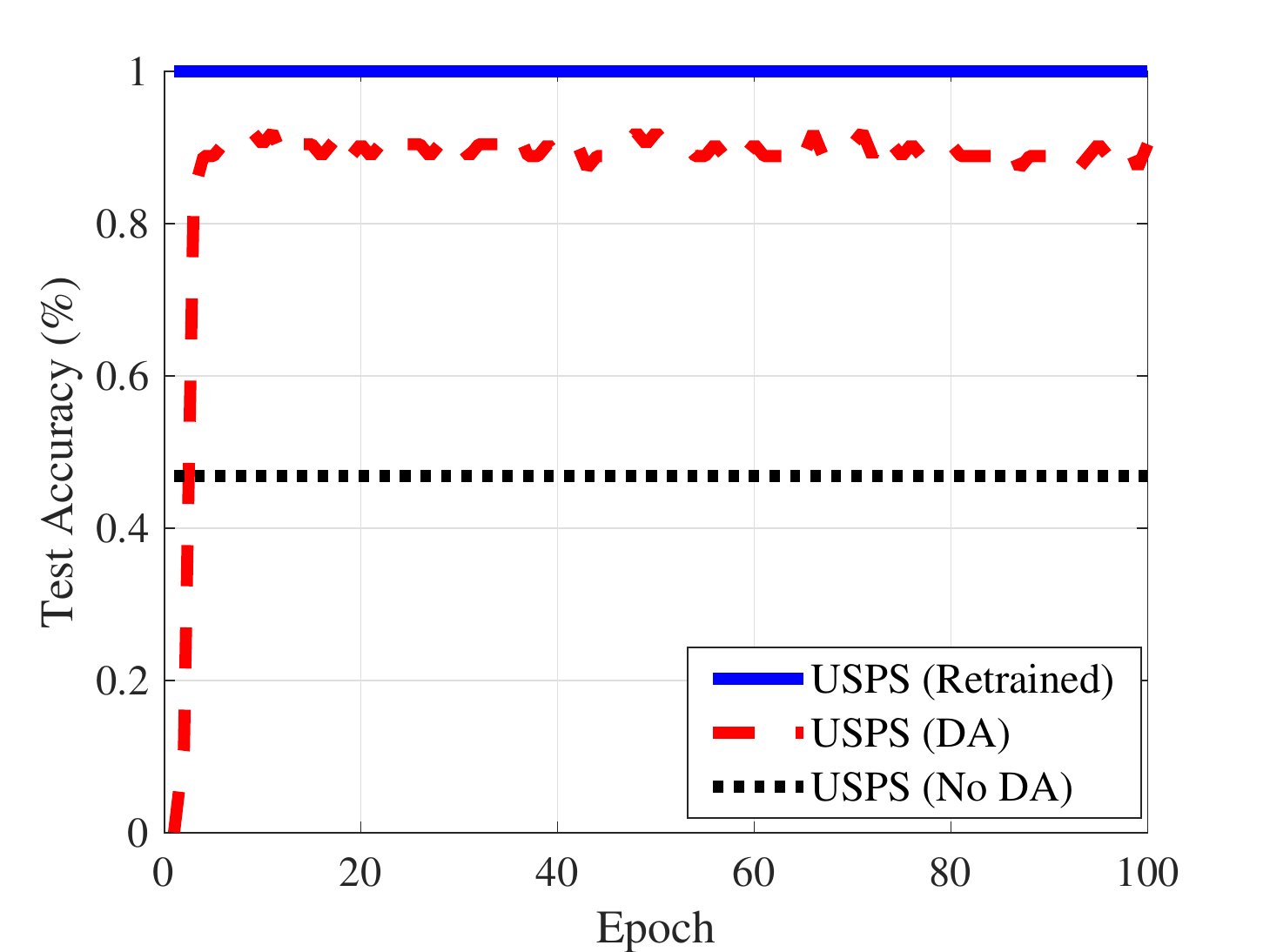}}
% \vspace{-10pt}
\subfigure[Accuracy against $CR$.]{
\label{fig:subfig:usps_cr}
\includegraphics[scale=0.5]{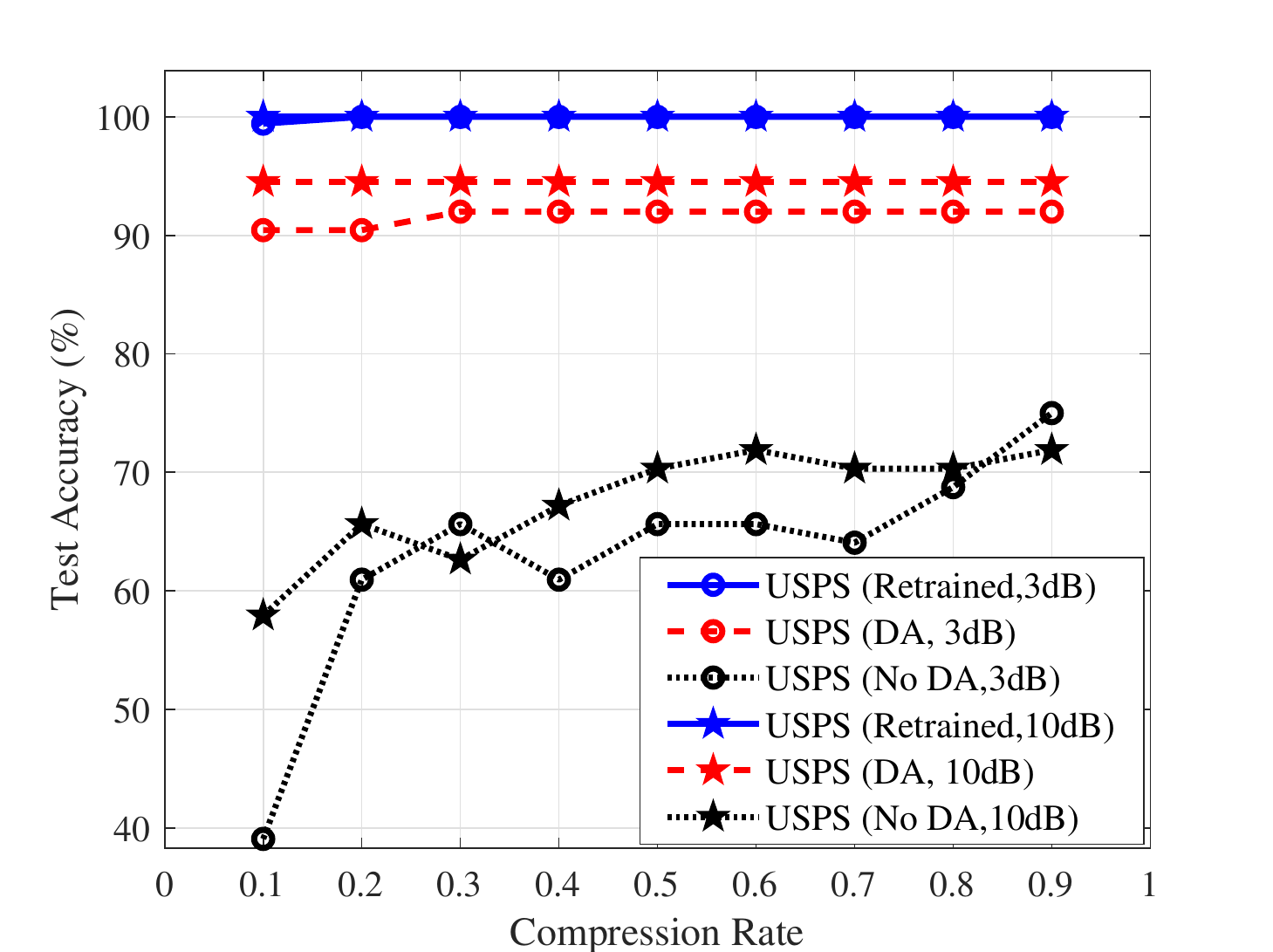}}

\caption{The accuracy of digit recognition with DA from USPS to MNIST.}
% \vspace{-15pt}
\label{fig:transfer_mnist_USPS}
\end{figure}

\begin{figure}[tb]
\centering
\subfigure[Up/down-sampled.]{
\includegraphics[scale=1.3]{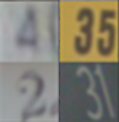}}
\subfigure[After DA.]{
\includegraphics[scale=1.3]{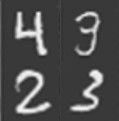}}
\subfigure[Recovered.]{
\includegraphics[scale=1.3]{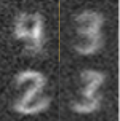}}
\caption{Examples of semantic communication for digit recognition with the SVHN dataset. }
\label{fig:DA_MNIST} %% the pragmatic outputsfor entire figure
\vspace{-12pt}
\end{figure}

\begin{figure}[tb]
\centering
\subfigure[Up/down-sampled.]{
\includegraphics[scale=1.47]{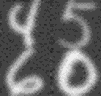}}
\subfigure[After DA.]{
\includegraphics[scale=1.47]{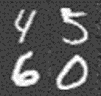}}
\subfigure[Recovered.]{
\includegraphics[scale=1.47]{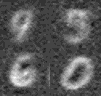}}
\caption{Examples of semantic communication for digit recognition with the USPS dataset. }
\label{fig:DA_MNIST2} %% the pragmatic outputsfor entire figure
\vspace{-12pt}
\end{figure}

This subsection is meant to show that a well trained semantic coding network can be re-used for different observed datasets. For example, here the semantic coding network is trained with the MNIST dataset as the library set, while the SVHN dataset and the USPS dataset are picked as the observed dataset respectively.

Fig. \ref{fig:transfer_mnist_SVHN} shows the accuracy performance of the digit recognition task for all three methods when SVHN is the observed dataset. Here we set $\textit{CR}=0.1$ and SNR$=3$dB. Fig. \ref{fig:subfig:svhn_epoch} plots the accuracy performance against the number of training epochs. It shows that our method can converge to the ideal result almost within 80 epochs. Fig. \ref{fig:subfig:svhn_cr} plots the accuracy performance against compression rate. We can see that the accuracy performance of our method is very close to that of the ideal case and far better than that of the method without DA.

Fig. \ref{fig:transfer_mnist_USPS} shows the accuracy performance of the digit recognition task when USPS is the actually observed dataset. Here, we still set $\textit{CR}=0.1$ and SNR$=3$dB. Fig. \ref{fig:subfig:usps_epoch} plots the accuracy performance against the number of training epochs. In this case, it shows that our method can converge to the ideal result almost within 5 epochs, and the difference in classification accuracy can be less than $10\%$. 
Meanwhile, Fig. \ref{fig:subfig:usps_cr} plots the accuracy performance against the compression rate. Since the USPS dataset is more similar to MNIST than SVHN, we can see that our method performs even better on the USPS dataset than on the SVHN dataset. As shown in the results of this section, the DA network in our system converges with only a few epochs and dozens of training samples, especially when two datasets share a high similarity.  

The visible results of the experiments above are shown in Fig. \ref{fig:DA_MNIST} and Fig. \ref{fig:DA_MNIST2}. We remain our focus on the case where $\textit{CR}=0.1$ and SNR$=3$dB. We can see that, for both experiments, the data adaptation network can convert the actually observed images into corresponding library images, while the semantic information is successfully kept. Under such a low $\textit{CR}$, the recovered images are blurred for human eyes, but still can be recognized by computers.

As a reference, the PAD, as introduced in Section \ref{sec:analysis}, between the USPS dataset and the MNIST dataset is 0.5, while the PAD between the SVHN dataset and the MNIST dataset is 1.48. It means that the similarity between the MNIST dataset and the USPS dataset is larger than the similarity between the MNIST dataset and the SVHN dataset. These numbers coincide with the visible results in Fig. \ref{fig:DA_MNIST} and Fig. \ref{fig:DA_MNIST2}, and also coincide with the performances shown in Fig. \ref{fig:transfer_mnist_SVHN} and Fig. \ref{fig:transfer_mnist_USPS}.\par

\subsubsection{DA-Based Semantic Communication for CIFAR10}

\begin{figure}[tb]
\centering
\subfigure[Accuracy against epoch.]{
\label{fig:subfig:stl_epoch} 
\includegraphics[scale=0.5]{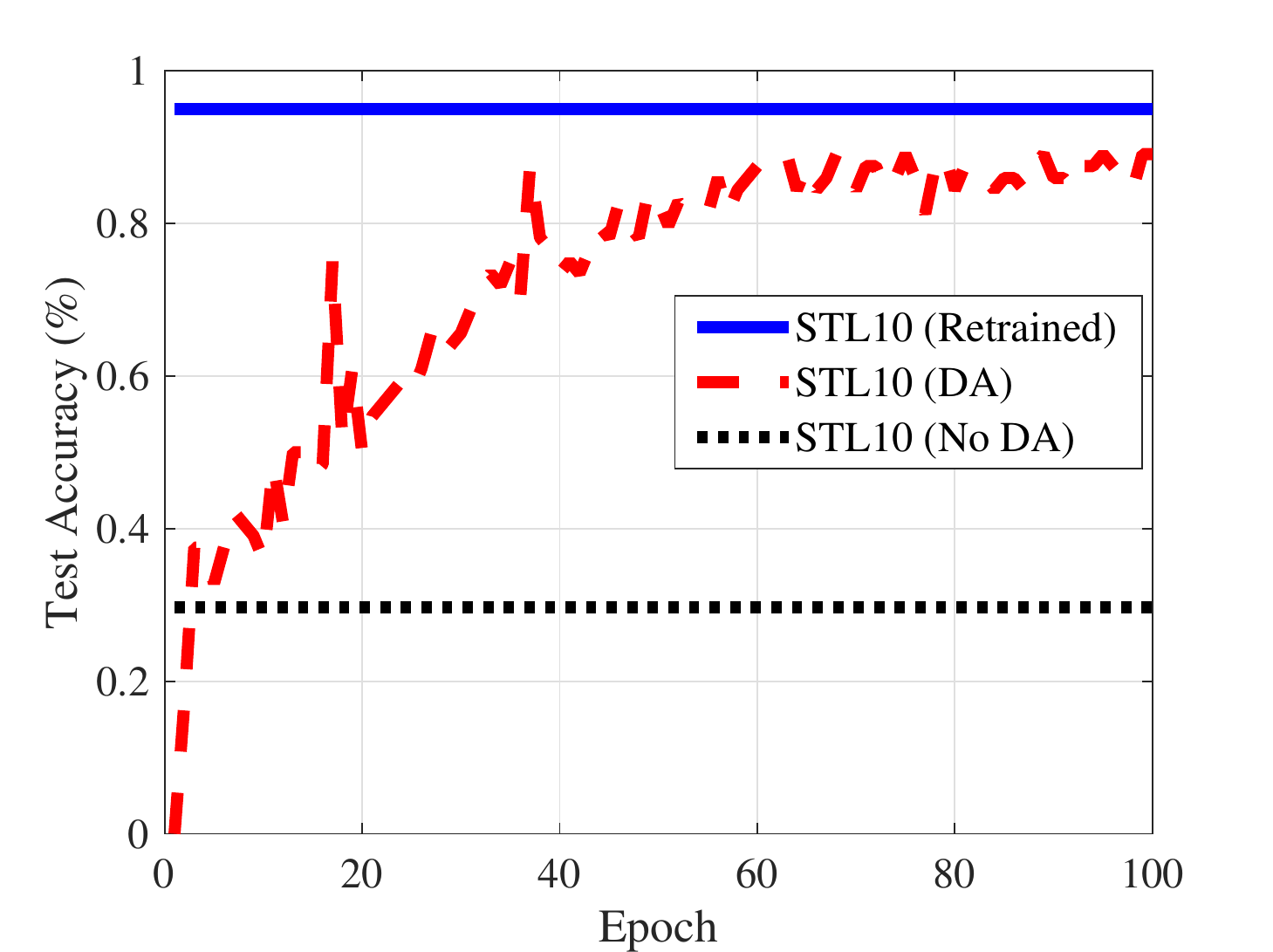}}
\subfigure[Accuracy against $CR$.]{
\label{fig:subfig:stl_cr} 
\includegraphics[scale=0.5]{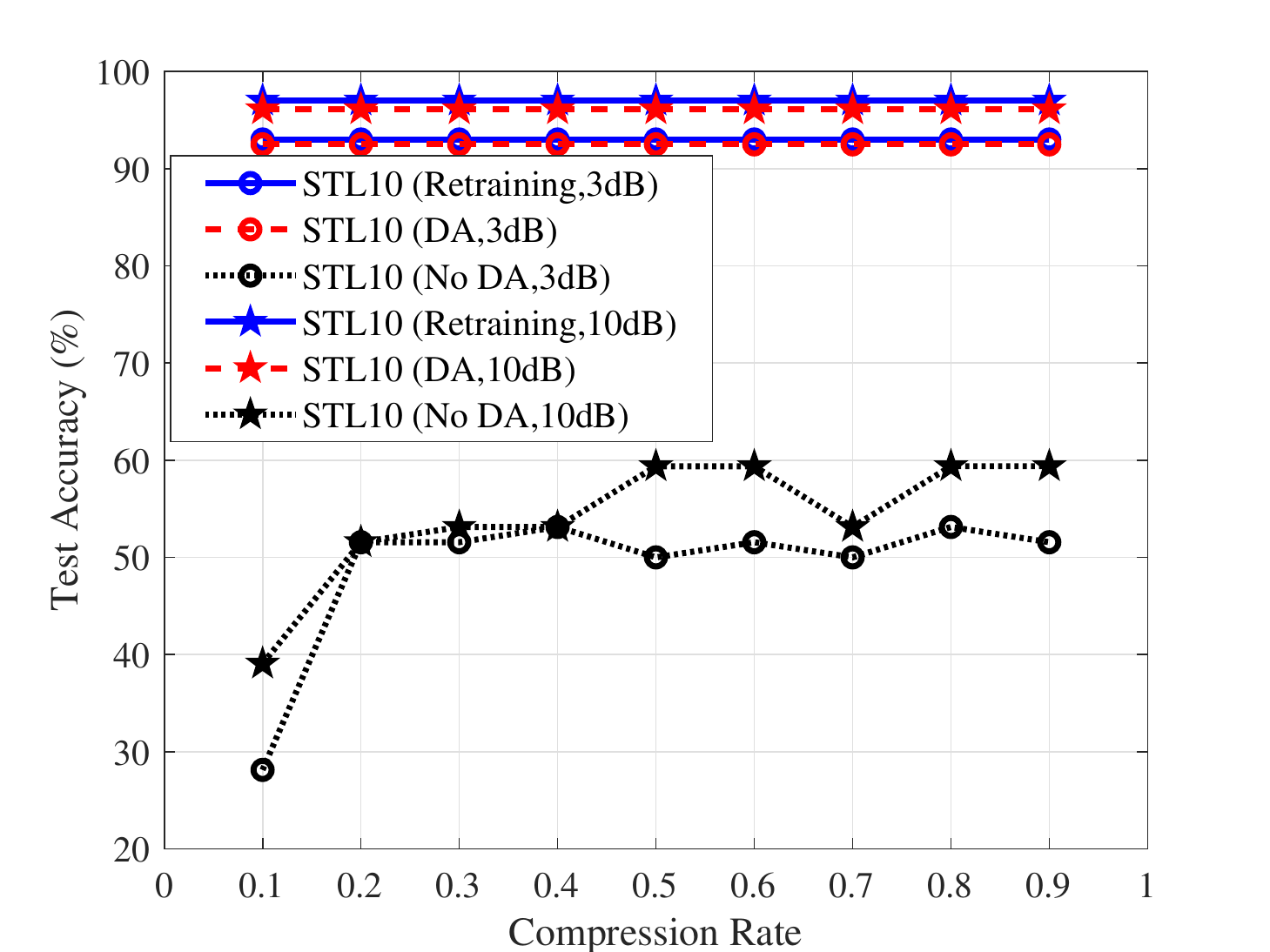}}
\caption{The accuracy of image classification with DA from STL10 to CIFAR10.}
\vspace{-12pt}         
\label{fig:transfer_cifar}
\end{figure}

\begin{figure}[t]
\centering
\subfigure[Up/down-sampled.]{
\includegraphics[scale=0.9]{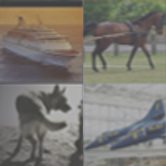}}
\subfigure[After DA.]{
\includegraphics[scale=0.9]{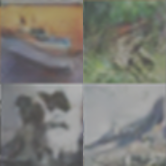}}
\subfigure[Recovered.]{
\includegraphics[scale=0.9]{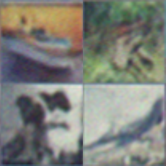}}
\caption{Examples of semantic communication for image classification with the STL10 dataset.}
\vspace{-15pt}           
\label{fig:subfig:STL2CIFAR}
\end{figure}

This experiment is meant to show that our method can be applied to different combinations of the library dataset and the observed dataset. As mentioned, in this experiment, we take the library set as the CIFAR-10 dataset and the observed set as the STL-10 dataset. The contents of images are also extended from digits to multiple objects, including animals, ships, airplanes, etc.

Fig. \ref{fig:subfig:stl_epoch} gives the classification accuracy performance for all three methods under different numbers of training epochs, where $\textit{CR}=0.1$ and SNR$=3$dB. As can be seen, with less than 80 epochs, the classification accuracy of our method is only $5\%$ less than that of the re-training schemes. The final result of the accuracy performance with our DA method is $59.3\%$ higher than that of the scheme without DA. Meanwhile, Fig. \ref{fig:subfig:stl_cr} shows the accuracy under different compression rates. It can be seen that our proposed DA method always outperforms the scheme without DA, and can be as good as the ideal case, for all $\textit{CR}\in[0.1,0.9]$.

Fig. \ref{fig:subfig:STL2CIFAR} presents the visualized results. As can be seen, the semantic information on the data after DA remains unchanged. Though images after DA are more blur than the raw images, they still can keep the outlines of the objects, and hence keeps the semantic information for classifying these images. 
\vspace{-8pt}

\section{Conclusion} \label{sec:conclusion}

In this paper, we proposed an NN-based semantic communication system, where the transmitter is unaware of the specific task at the receiver and the dataset required by the receiver is dynamic. This system contains two individual parts, namely the semantic coding network and the data adaptation network. To address the task-unaware issue, we establish a receiver-leading training process, which can jointly train the encoder and decoder in the semantic coding network, without leaking direct information about the pragmatic task to the transmitter. Based on the proposed training process, a semantic-oriented loss function is proposed, which can carefully choose the proper information distortion metric for different pragmatic tasks. To deal with the dynamic dataset, we introduce a domain adaptation-based method to save communication costs for re-training the system. Numerical results show that our semantic coding network can enhance the performance of the pragmatic task by losing very little bit-wise performance, and our data adaptation network can have an upper bound achieving performance over different combinations of observed datasets and library datasets.

This paper is the first work addressing the applicability of semantic or task-oriented communication in practical scenarios with unequal participants and varying data environments. There are many interesting problems to further investigate in the future, such as how to coordinate the training of semantic encoder and decoder with minimum communication cost, how to adapt the semantic communication to a time-varying channel environment, to name a few. We believe such studies can accelerate the deployment of semantic communications in practical scenarios, like on IoT devices or UAVs.

\ifCLASSOPTIONcaptionsoff
  \newpage
\fi

% % \nocite{*}
\bibliographystyle{IEEEtran}
% % \bibitem{ref1}
% % \bibliographystyle{unsrt}
\bibliography{refer.bib}

\end{document}